\documentclass[11pt]{article}
\pdfoutput=1
\usepackage[margin=1.0in]{geometry}
\usepackage{amsmath,amssymb,graphicx}
\usepackage{hyperref}
\usepackage{slashed}
\usepackage{bm}

\usepackage[T1]{fontenc}

\usepackage[utf8]{inputenc}

\usepackage{color,soul}

\usepackage[greek,english]{babel}
\newcommand{\cpi}{\text{\greektext p}}

\usepackage[font=small,labelfont=bf]{caption}

\usepackage{tabu}

\newcommand{\iu}{{\mathrm i}}
\newcommand{\E}{{\mathrm e}}

\newcommand{\dd}{{\mathrm{d}}}

\newcommand{\omR}{\omega_\mathrm{R}}
\newcommand{\omI}{\omega_\mathrm{I}}

\newcommand{\df}{\mathrel{:=}}

\newcommand{\be}{\begin{equation}}
\newcommand{\ee}{\end{equation}}

\usepackage{xcolor}

\title{\bf Quasinormal Modes of Charged Fields in Reissner--Nordstr{\"o}m Backgrounds by Borel--Pad{\'e} Summation of Bender--Wu Series}
\author{Dan Stefan Eniceicu and Matthew Reece\\
{\small \color{gray} \texttt{eniceicu~(@college.harvard.edu), mreece~(@g.harvard.edu)}}\\
{\small Department of Physics, Harvard University, Cambridge, MA, 02138}}

\begin{document}
\maketitle

\begin{abstract}
We extend recent work of Hatsuda on the computation of quasinormal mode frequencies via analytic continuation of bound state energies and Borel--Pad{\'e} resummation of the Bender--Wu perturbation series to the case of charged fields in the background of Reissner--Nordstr{\"o}m black holes. We compare the quasinormal mode frequencies obtained in this manner to calculations using Leaver's method of continued fractions, and find good agreement for damped modes (DMs) with imaginary part remaining finite in the extremal limit. We also present numerical evidence that the frequencies of certain zero-damped modes (ZDMs) with imaginary part tending to zero in the extremal limit can be computed when constructing the Bender--Wu expansion about a peak of the potential {\em inside} the outer horizon of the black hole. 
\end{abstract}


\section{Introduction}
\label{sec:introduction}

Black hole solutions have a characteristic spectrum of decaying perturbations known as {\em quasinormal modes}. For black holes in asymptotically flat space, these are defined by the boundary condition that waves are not emerging from the black hole horizon or propagating in from spatial infinity. Quasinormal modes are of interest for real-world observations, as they describe the ``ringdown'' phase of black hole mergers observed by gravitational wave interferometers like LIGO and Virgo. They are also of theoretical interest, providing a probe of black hole properties that may shed light on quantum gravity. Readers seeking a general introduction to the topic (and references to the earliest literature) can consult one of the thorough review articles \cite{Nollert:1999ji, Kokkotas:1999bd, Berti:2009kk, Konoplya:2011qq} on the subject, or the classic monograph by Chandrasekhar \cite{Chandrasekhar:1985kt}.

Although the problem of finding the quasinormal mode spectrum of a black hole is superficially similar to the more familiar problem of computing bound state (normal mode) frequencies (e.g., of a quantum mechanical particle in a potential), it tends to be more difficult. One source of difficulty is that quasinormal mode wavefunctions grow exponentially at spatial boundaries. We work in the sign convention that the time dependence of a mode is $\E^{-\iu \omega t}$. In this case, modes that decay exponentially in time have ${\rm Im}\, \omega < 0$. We denote the real and imaginary parts of a quasinormal mode frequency as
\begin{equation}
\omega = \omR - \iu \omI.  
   \label{eq:omRomI}
\end{equation}
The size of $\omI > 0$ determines how quickly a quasinormal mode decays. We parametrize the radial coordinate outside a black hole with a tortoise coordinate $y$ that goes to $-\infty$ at the horizon and $+\infty$ at spatial infinity. (Another common notation for this coordinate is $r_*$.) The boundary condition at a black hole horizon is that the mode is entering the horizon. That is to say, $y$ is becoming more negative as time advances, i.e., the  mode behaves as $\exp[-\iu \omega (t + y)]$ as  $y \to -\infty$. This implies that the mode grows exponentially toward the horizon, as $\exp(\omI |y|)$. Similarly, the boundary condition at spatial infinity is that the mode is outgoing, i.e., behaving as $\exp[-\iu \omega(t - y)]$ as $y \to +\infty$, which grows exponentially as $\exp(\omI y)$. Because the wavefunctions grow exponentially at the  boundaries, simply running a standard ODE solver and imposing the boundary conditions numerically will typically fail, as it would require computing the solution with exponentially high accuracy. Many clever numerical and semi-analytic schemes have been developed over the years to circumvent this problem. One example is Leaver's method, which begins with an ansatz that factors out the asymptotic dependence of the solution as $y \to \pm \infty$ and then solves for the remaining unknown factor through a recursion relation. This recursion relation can be cast in the form of a continued fraction expansion, which allows for efficient numerical computation \cite{Leaver:1985ax, Leaver:1990zz}. Another approach, the asymptotic iteration method, uses an invariance of the right-hand side of a particular form of the differential equation under repeated differentiation to construct an accurate approximation to the solution \cite{Cho:2009cj, Cho:2011sf}. Another method is based on the WKB approximation \cite{Schutz:1985zz,Iyer:1986np, Zaslavsky:1991ug}; recently, Pad\'{e} resummation has been used to improve this method \cite{Matyjasek:2017psv,  Konoplya:2019hlu}.  In some cases, there is substantial overhead in preparing analytic calculations to adapt a numerical scheme to each new quasinormal mode problem.

A recent algorithm developed by Hatsuda \cite{Hatsuda:2019eoj} relies on Borel--Pad{\'e} summation of high-order perturbative results. The high orders of perturbation theory are computed with the \texttt{BenderWu} software package \cite{Sulejmanpasic:2016fwr}, which extends the classic results of Bender and Wu on the anharmonic oscillator \cite{Bender:1990pd} to generic potentials. Subsequent to Hatsuda's work, a closely related study has appeared in \cite{Matyjasek:2019eeu}. Hatsuda's algorithm is quite flexible, as it can be immediately applied to any spherically-symmetric quasinormal mode problem that can be cast in the form of the radial master equation
\begin{equation}
\left(\frac{\dd^2}{\dd y^2} + \omega^2 - V(y)\right) \phi(y) = 0.
   \label{eq:master1}
\end{equation}
In this paper, we aim to extend this algorithm in new directions. Even with the assumption of spherical symmetry, not all quasinormal mode problems take the form \eqref{eq:master1}. The quasinormal modes of charged fields in the background of a charged black hole obey a more general radial master equation,
\begin{equation}
\left(\frac{\dd^2}{\dd y^2} + \left[\omega - q K(y)\right]^2 - V(y)\right) \phi(y) = 0,
   \label{eq:master2}
\end{equation}
where the function $K(y)$ is the radial profile of the gauge field in the background. In this paper, we generalize Hatsuda's algorithm to apply to problems of the form \eqref{eq:master2}, and demonstrate that it can accurately compute quasinormal mode spectra. This provides a flexible approach that can be straightforwardly applied to any problem of the form \eqref{eq:master2}.

Another area in which we can extend Hatsuda's results relates to the existence of what are known as ``zero-damped modes'' (ZDMs), which seem to be missed in a first attempt at applying this algorithm. We will explore ways that the algorithm might be used to search for ZDMs by perturbing around more general extrema of the potential. Because ZDMs have received attention in the literature only relatively recently, we will now briefly review them. For a generic black hole that is not close to extremality, its temperature is set by the inverse of its radius, and one expects quasinormal mode frequencies to be of order this temperature. As charged and/or spinning black holes approach extremality, their temperature goes to zero while their radius remains finite. In this case, quasinormal modes can belong to two families: damped modes (DMs) for which $\omI$ remains finite and nonzero (of order the inverse black hole radius) in the extremal limit, and zero-damped modes for which $\omI \to 0$ in the extremal limit (linearly with the temperature, in known examples). Much of the literature on quasinormal modes has focused on DMs, whereas ZDMs have sometimes been overlooked. They may be missed by some numerical approaches to computing quasinormal modes. On the other hand, ZDMs are often tractable to study analytically in the near-extremal limit. The family of ZDMs can be computed even away from extremality. A crossover appears at a critical fraction of extremality when the mode of smallest $\omI$ switches from a DM to a ZDM (see, e.g., \cite{Richartz:2014jla}). 

The first indication that ZDMs exist came from an analytic approximation by Detweiler to the Teukolsky equation for modes of nearly-extremal Kerr black holes \cite{Detweiler:1980gk} (building on \cite{Starobinskil:1974nkd, Teukolsky:1974yv}). Additional studies followed \cite{Sasaki:1989ca, Andersson:1999wj}. Eventually, a clear numerical picture has emerged, demonstrating that both DMs and ZDMs exist not only for Kerr black holes but for Reissner--Nordstr{\"o}m and Kerr--Newman black holes, and for both the gravitational and electromagnetic perturbations of these black holes and for external fields in the black hole background \cite{Yang:2012pj, Yang:2013uba, Konoplya:2013rxa, Richartz:2014jla, Cook:2014cta, Dias:2015wqa, Zimmerman:2015trm}. This numerical picture bolsters an analytic argument in the near-extremal limit developed in a series of papers \cite{Gruzinov:2007ai, Hod:2007tb, Hod:2008zz, Hod:2010hw}. The existence of the ZDMs is important for a conjectured ``universal relaxation bound,'' which postulates that for any black hole of temperature $T_{\rm BH}$ there must exist {\em some} quasinormal mode (of some field, not necessarily a gravitational perturbation) for which $\omI \lesssim \cpi T_{\rm BH}$ \cite{Hod:2006jw}.

The outline of this paper is as follows. In \S\ref{sec:prelim}, we review Hatsuda's algorithm and the Bender--Wu series that provides a crucial input to the algorithm. We explain why this approach is not sufficiently general to calculate the quasinormal modes of charged fields. In the process, we review the Reissner--Nordstr{\"o}m solution and the form of the quasinormal mode equations for charged fields. This section is largely a review and much of it can be skipped by informed readers. In \S\ref{sec:algorithm}, we generalize the Bender--Wu series to the class of differential equations that arise for charged fields. This constitutes the main result of our paper. In \S\ref{sec:validateDMs}, we validate our proposed algorithm by comparing calculations of damped quasinormal modes using our generalization  of Hatsuda's algorithm to calculations using the classic method of Leaver \cite{Leaver:1985ax}, finding good agreement. The algorithm requires a choice of a critical radius $r_0$ to expand around, which for uncharged fields is simply an extremum of $V(r)$, but more generally is a solution to \eqref{eq:r0extrema}. In \S\ref{sec:ZDMs}, we present some preliminary exploration of what happens when we make alternative choices of $r_0$. We find that for a critical scalar field mass, a bifurcation in the solutions for $r_0$ occurs, and near this bifurcation point our numerical method breaks down. On the other hand, we find indications that if we choose $r_0$ to lie inside the black hole's outer horizon, Hatsuda's method can be used to compute zero-damped modes. We emphasize that this is an empirical result. We have not analytically understood why the solution to the equation inside the black hole's outer horizon obeys the appropriate quasinormal mode boundary conditions at $r \to \infty$. Finally, we conclude in \S\ref{sec:outlook} by highlighting some important questions that will be the focus of future work.

\section{Preliminaries: reviewing Hatsuda's method and charged quasinormal mode equations}
\label{sec:prelim}

In this section, we will provide a sketch of the ingredients of Hatsuda's method for the calculation of quasinormal modes \cite{Hatsuda:2019eoj}. We will then explain two areas in which we can extend Hatsuda's method: first, to compute quasinormal modes for charged fields in a charged black hole background, for which the quasinormal mode equations do not take the appropriate form to be solved in this manner; second, to explore the zero-damped modes that are known to exist in the near-extremal limit.

\subsection{The Hatsuda algorithm}
\label{subsec:Hatsudaalgorithm}

The strategy of the calculation is as follows. The radial equation for quasinormal modes can be written in terms of the ``tortoise coordinate'' $y$,
\begin{equation}
\dd y = \frac{\dd r}{f(r)},
\end{equation}
where $f(r)$ is the prefactor  of $\dd t^2$ in the metric which vanishes at the horizon. The coordinate $y$ runs from $-\infty$ at the horizon to $+\infty$ at spatial infinity. In terms of this coordinate, the radial equation for typical gravitational and electromagnetic perturbations of black holes takes the form of \eqref{eq:master1}. The potential $V(y)$ typically has a maximum outside the horizon. Hatsuda's observation, building on similar past work on analytic continuation \cite{blome1984quasi, Ferrari:1984ozr, Ferrari:1984zz}, is that the problem with a formal expansion parameter $\epsilon$,
\begin{equation}
\left(\epsilon^2 \frac{\dd^2}{\dd y^2} + \omega^2 - V(y)\right) \phi(y) = 0,
   \label{eq:masterepsilon}
\end{equation}
can  be viewed as an analytic continuation of a time-independent Schr\"odinger equation,
\begin{equation}
\left(- \hbar^2 \frac{\dd^2}{\dd y^2} - E - V(y)\right) \phi(y) = 0,
   \label{eq:schrodinger}
\end{equation}
which has a potential energy $-V(y)$. This inverted potential has a {\em minimum} and so we expect conventional quantum mechanical bound states with wave functions decaying exponentially at $y \to \pm \infty$. Upon analytic continuation $\hbar \mapsto \iu \epsilon$, the bound state energies $E_n$ become the quasinormal mode squared frequencies $\omega_n^2$,
\begin{equation}
\omega_n^2\Big|_{\epsilon = 1} = - E_n\Big|_{\hbar = \iu}.
\end{equation}
The exponential decay of the bound state wavefunctions analytically continues to the expected exponential behavior of the quasinormal mode wavefunctions  at the horizon and at spatial infinity, as we discussed in \S\ref{sec:introduction}.

To obtain the analytic continuation, Hatsuda uses the existence of an efficient calculation technique for the series  expansion of the  bound state energies in the  formal expansion parameter $\hbar$, namely the Bender--Wu recursive calculation about the quadratic minimum of a potential, which we review in the next subsection. The series expansion,
\begin{equation}
\epsilon_n \df \frac{E_n-V_0}{2\hbar} \approx \sum_{k=0}^\infty \epsilon_n^{(k)} \hbar^k,
\end{equation}
is an asymptotic series. This series can be resummed by Borel--Pad\'e summation. One first computes the Borel transformed series,
\begin{equation}
{\cal B}[\epsilon_n](\zeta) \df \sum_{k=0}^\infty \frac{1}{k!}  \epsilon_n^{(k)} \hbar^k,
\end{equation}
which has better convergence properties than the original asymptotic series. Hatsuda found empirically that the Borel series can be summed, and that an efficient approximation is obtained by replacing the Borel series with a high-order $[M/N]$ Pad\'e approximant and then computing the inverse Borel transform (Laplace transform) of the Pad\'e-approximated Borel series. In \cite{Hatsuda:2019eoj}, this Borel--Pad\'e summation procedure was checked to agree to many decimal places with calculations obtained by Leaver's method \cite{Leaver:1985ax} for gravitational perturbations of Schwarzschild and odd-parity perturbations of Reissner--Nordstr{\"o}m black holes.

\subsection{The Bender--Wu series calculation}

An important step in Hatsuda's algorithm is the computation of the perturbative approximation to energy eigenvalues as a series  in the formal parameter $\hbar$, to high orders. This is performed with the \texttt{BenderWu} Mathematica package \cite{Sulejmanpasic:2016fwr} of Sulejmanpasic and {\"U}nsal. In this subsection, we briefly review the Bender--Wu algorithm.

The Bender--Wu algorithm begins with a quadratic minimum of the potential, and assumes that the potential and its energy eigenvalues can be written as a formal power series in a parameter $g$, expanded around the harmonic oscillator solutions in this quadratic potential. That is, we solve the  Schr{\"o}dinger equation
\begin{equation}
-\frac{1}{2} \psi''(x) + V(x)  \psi(x) = E \psi(x),
\end{equation}
where
\begin{equation}
V(x) = \sum_{n = 0}^\infty V_n g^n x^{n+2} \quad \text{and} \quad E =  \sum_{n=0}^\infty E_n g^n,
\end{equation}
assuming without loss of generality that the quadratic minimum  is at $x = 0$. The wavefunction is taken to be the harmonic oscillator ground state wavefunction multiplied by an unknown function $Y(x)$, which again is taken as a power series in $g$ whose coefficients are unknown functions of $x$:
\begin{equation}
\psi(x) = Y(x) \E^{-V_0 x^2}, \quad Y(x) = \sum_{n = 0}^\infty Y_n(x) g^n.
\end{equation}
In the $n = 0$ case, the function $Y_0(x)$ turns out to be one of the Hermite polynomials, $H_\nu(x)$, and the  corresponding eigenvalue is $E_0 = \sqrt{2 V_0} (\nu + 1/2)$, the standard harmonic oscillator result.

The achievement of the Bender--Wu technique is to systematically work out higher orders in  $n$. Substituting the ans{\"a}tze into the Schr\"odinger equation, one obtains recursion relations that relate the different functions $u_n(x)$. As a further step, the individual functions $Y_n(x)$ are expanded in powers of $x$. Importantly, one can argue that there is a maximum power $x^{K_{n,\nu}}$ that one must consider. It depends on both the order $n$ in perturbation theory to which we are working and the level $\nu$ of the bound state we consider. Then we can take
\begin{equation}
Y_n(x) = \sum_{k=0}^{K_{n,\nu}} A^k_n x^k.
\end{equation}
Upon making this substitution in our ansatz for the wavefunction, we obtain a complicated but solvable recursion for the unknown values $A^k_n$ and $E_n$ in terms of the coefficients $V_n$.

We will omit further details in this section because we will shortly generalize this procedure to a larger set of differential equations in \S\ref{sec:algorithm}. 

\subsection{The Reissner--Nordstr{\"o}m black hole: a mini-review}
\label{sec:minireview}

Our primary interest in this paper is in the quasinormal modes of charged scalar fields in a Reissner--Nordstr{\"o}m background. We first review this solution, in part to fix our conventions. The action for this theory (we work in a mostly-plus signature) is
\begin{equation}
S = \int \dd^Dx\,\sqrt{-g} \left[\frac{1}{2\kappa^2} {\cal R} - \frac{1}{4} F_{\mu \nu}F^{\mu \nu} - g^{\mu  \nu} (D_\mu \phi)^* D_\nu \phi - \mu^2 \phi^* \phi \right],
\end{equation}
with $\cal R$ the Ricci scalar, $F_{\mu \nu} = \partial_\mu A_\nu - \partial_\nu A_\mu$ the gauge field strength, and $D_\mu \phi \df \nabla_\mu \phi - \iu q A_\mu \phi$ the gauge covariant derivative of the charged field $\phi$. We can also write  the gravitational coupling $\kappa$ in terms of Newton's constant $G_N$, $\kappa^2 = 8 \cpi G_N$.

The Reissner--Nordstr{\"o}m black hole solution is given in terms of two parameters $r_S$ and $r_Q$ by
\begin{align}
\dd s^2 &= -F(r) {\rm d}t^2 + \frac{1}{F(r)} {\rm d}r^2 + r^2 {\rm d}\Omega_{D-2}^2, \nonumber \\
F(r) &= 1 - \left(\frac{r_S}{r}\right)^{D-3} + \left(\frac{r_Q}{r}\right)^{2(D-3)},  \nonumber \\
F_{tr} &= -\frac{Q}{\Omega_{D-2} r^{D-2}} \quad \text{where} \quad \Omega_{D-2} \df \frac{2\cpi^{(D-1)/2}}{\Gamma\left(\frac{D-1}{2}\right)}.
\label{eq:RNsolution}
\end{align}
The vector potential $A_t$ is obtained by integrating $F_{tr}$, up to a constant gauge ambiguity $K_\infty$:
\begin{equation}
A_t(r) = - \frac{Q}{(D-3) \Omega_{D-2}} \frac{1}{r^{D-3}} - K_\infty.
 \label{eq:RNvectorpotential}
\end{equation}
The charge $Q$ appearing in the field strength $F_{tr}$ and the mass $M$ of the black hole are related to the length scales $r_S$ and $r_Q$:
\begin{align}
M &= \frac{(D-2) \Omega_{D-2}}{2\kappa^2} r_S^{D-3}, \nonumber \\
Q &= \frac{\sqrt{(D-2)(D-3)}\Omega_{D-2}}{\kappa} r_Q^{D-3}.
\label{eq:RNMQ}
\end{align}
The inner and outer horizons of the black hole are located at the coordinates $r_-$ and $r_+$ where $F(r_\pm) = 0$:
\begin{equation}
r_\pm^{D-3} = \frac{1}{2} \left(r_S^{D-3} \pm \sqrt{r_S^{2(D-3)} - 4 r_Q^{2(D-3)}}\right).
\end{equation}
We can write $F(r)$ in a factored form in terms of the horizon coordinates:
\begin{equation}
F(r) = \left[1 - \left(\frac{r_+}{r}\right)^{D-3}\right]\left[1 - \left(\frac{r_-}{r}\right)^{D-3}\right].
\end{equation}
The black hole has a temperature
\begin{equation}
T = \frac{(D-3)}{4\cpi r_+} \frac{r_+^{D-3} - r_-^{D-3}}{r_+^{D-3}},
\end{equation}
which is the surface gravity divided by $2\cpi$. The black holes obey an extremality bound
\begin{equation}
Q \leq Q_\text{ext} \df \sqrt{\frac{D-3}{D-2}} \kappa M.
\end{equation}
When this bound is saturated, $r_+ = r_-$. The temperature goes to zero in the extremal limit.

\subsection{Charged scalar quasinormal mode equations}

We consider the charged scalar as a probe field, that is, we assume that we are interested in small field values that do not significantly perturb the black hole background. This suffices to understand the spectrum of quasinormal modes. Rather than specializing to the Reissner--Nordstr{\"o}m background immediately, we can work out the equations of motion for a general spherically symmetric ansatz
\begin{align}
{\rm d}s^2 &= -F(r) {\rm d}t^2 + G(r) {\rm d}r^2 + H(r) {\rm d}\Omega_{D-2}^2, \nonumber \\
A_\mu &= -K(r) \delta^t_\mu.
\end{align}
We can search for scalar field solutions of the form
\begin{equation}
\phi(t, r, \theta_1, \ldots, \theta_{D-2}) = \E^{-\iu \omega t} \sum_L R_L(r) Y_L(\theta_1, \ldots, \theta_{D-2}),
\end{equation}
where $L = (l_1, \ldots, l_{D-2})$ is a tuple of indices with $|l_1| \leq l_2 \leq l_3 \leq \cdots \leq l_{D-2}$ parametrizing the generalized spherical harmonics, which are eigenfunctions of the Laplace--Beltrami operator on $S^{D-2}$. The corresponding eigenvalue is $-l_{D-2}(l_{D-2}+D-3)$. If we define
\begin{equation}
\chi(r) \df \frac{1}{2} \log \left(\frac{F(r)}{G(r)} H^{D-2}(r)\right),
\end{equation}
then we find that the radial wavefunction $R_L(r)$ obeys the equation
\begin{equation}
R_L''(r)+\chi'(r)R_L'(r)+G(r)\Bigg\{\frac{1}{F(r)}\big[\omega-qK(r)\big]^2-\left[\mu^2 + \frac{l_{D-2}\big(l_{D-2}+D-3\big)}{H(r)}\right]\Bigg\}R_L(r)=0.
\label{eq:chargedradialequation}
\end{equation}

To rewrite this in a more Schr{\"o}dinger-like form, we can define a generalized tortoise coordinate $y$ via 
\begin{equation}
\frac{\dd r}{\dd y}=\sqrt{\frac{F(r)}{G(r)}} \quad \Rightarrow \quad \frac{\dd^2r}{\dd y^2}=\frac{F(r)}{2G(r)}\left(\frac{F'(r)}{F(r)}-\frac{G'(r)}{G(r)}\right). \label{eq:tortoise}
\end{equation}
Rescaling our radial wavefunction by an $H$-dependent factor,
\begin{equation}
\Psi_L(r) = H(r)^{\frac{D-2}{4}} R_L(r),
\end{equation}
we find that the new function obeys the simple differential equation
\begin{equation}
\left(\frac{\dd^2}{\dd y^2} + \left[\omega - q K(r)\right]^2 - V(r)\right) \Psi_L(r) = 0,
\label{eq:chargedscalarQNMform}
\end{equation}
where $r$ is implicitly a function of $y$. The potential appearing in this equation can be computed from the metric via:
{\small 
\begin{equation}
V(r)=F(r)\Bigg\{\mu^2+\frac{l_{D-2}\big(l_{D-2}+D-3\big)}{H(r)}+\frac{D-2}{16G(r)}\left[(D-6)\frac{H'(r)^2}{H(r)^2}+4\frac{H''(r)}{H(r)}+2\frac{H'(r)}{H(r)}\left(\frac{F'(r)}{F(r)}-\frac{G'(r)}{G(r)}\right)\right]\Bigg\},
\label{eq:Vgeneral}
\end{equation}
}where primes denote derivatives with respect to $r$. Specializing to the Reissner--Nordstr{\"o}m solution where $G(r) = F(r)^{-1}$ and $H(r) = r^2$, this takes the simple form
\begin{equation}
V(r)=\frac{F(r)}{r^2}\left[\mu^2r^2+l_{D-2}\big(l_{D-2}+D-3\big)+\frac{(D-2)(D-4)}{4}F(r)+\frac{D-2}{2}rF'(r)\right],
\end{equation}
and the wavefunction $\Psi_L(r)$ is simply $r^{\frac{D-2}{2}} R_L(r)$.

\subsection{Opportunities to extend Hatsuda's algorithm}
\label{subsec:opportunities}

Finally, we have all of the ingredients to point out two areas where Hatsuda's algorithm can be extended. The first is that charged scalar quasinormal modes obey differential equations of the form \eqref{eq:chargedscalarQNMform}. When $K(r) = 0$, this can be analytically continued to a Schr{\"o}dinger equation where $\omega^2$ plays the role of the energy eigenvalue. However, in a charged black hole background where $K(r) \neq 0$, this is no longer the case: the equation does not have the form of an eigenvalue equation, because of the $\omega K(r)$ cross term. Hence, the pre-existing Bender--Wu calculation must be adapted to solve this more general differential equation.

\begin{figure}[!h]
\centering
\includegraphics[width=0.55\textwidth]{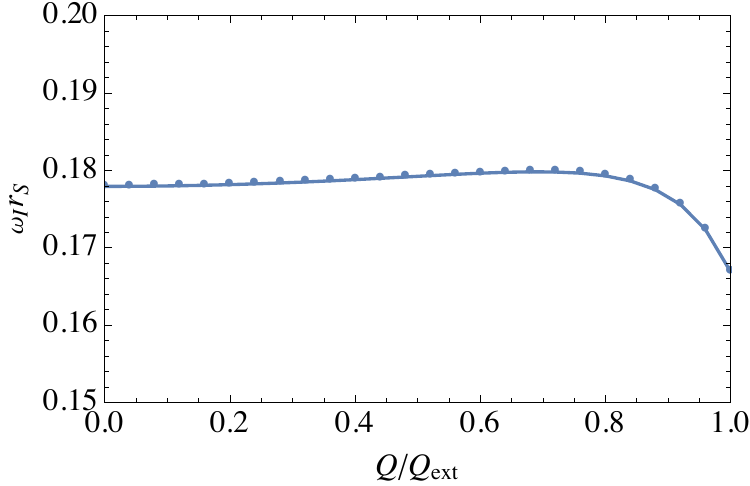}
\caption{Imaginary part $\omI$ of the  quasinormal mode frequency for odd-parity gravitational/electromagnetic perturbations of the $D = 4$  Reissner--Nordstr{\"o}m black hole, as computed for the $l=2, \nu = 0$ mode found with Hatsuda's method \cite{Hatsuda:2019eoj}. The horizontal axis shows the black hole charge measured as a fraction of extremality, whereas the vertical axis shows the frequency measured in units of $r_S^{-1} = (2G_N M)^{-1}$. The fact that $\omI$ remains finite as $Q \to Q_\mathrm{ext}$ indicates that the algorithm is finding a damped mode, not a zero-damped mode, in the extremal limit.}
\label{fig:HatsudaRNdamped}
\end{figure} 

The second place where Hatsuda's method could be extended is that the simplest implementation of the algorithm finds only DMs and not ZDMs. We illustrate this in Fig.~\ref{fig:HatsudaRNdamped} by showing the $l = 2, \nu = 0$ quasinormal mode frequency for the mixed gravitational/electromagnetic perturbations of the $D = 4$ Reissner--Nordstr{\"o}m black hole as computed with Hatsuda's method. We see that $\omI$ has only modest dependence on the black hole charge $Q$; in particular, although it becomes slightly smaller (when measured in units of $r_S^{-1}$) at $Q = Q_\mathrm{ext}$, it remains finite in this limit. This is the defining feature of a damped mode. We have also checked that several modes of higher $\nu$ computed by Hatsuda's method remain damped in the $Q \to Q_\mathrm{ext}$ limit. It has been argued that zero-damped modes exist for the Reissner--Nordstr{\"o}m black hole \cite{Zimmerman:2015trm}, which raises the question of how they might be found using an extension of Hatsuda's method.

\section{Algorithm}
\label{sec:algorithm}

In this section, we generalize the Bender--Wu recursive calculation to generate high-order perturbative approximations to solutions of the equation \eqref{eq:chargedscalarQNMform}. This method is the central result of our paper. Because the derivation is rather involved, we present a summary of the algorithm in \S\ref{subsec:summary} for readers who prefer to skip over the details. In \S\ref{sec:validateDMs}, we will present numerical results of the algorithm validated against calculations using Leaver's method.

\subsection{Extending the Bender--Wu Method for the Charged Scalar}

Our goal is to numerically solve for the quasinormal modes $\omega$ of equation \eqref{eq:chargedscalarQNMform} with the potential \eqref{eq:Vgeneral}, where $r$ is a function of $y$ defined implicitly via \eqref{eq:tortoise}. We will proceed by adapting the logic of the Bender--Wu analysis to this modified problem.
Let $r_0=r(y_0)$ be a point to be determined around which we expand $V(r)$. Let $y-y_0=gx$, and
\begin{align}
-V(r)&=\sum_{k=0}^\infty V_kg^kx^k, \nonumber \\
-qK(r)&=\sum_{k=0}^\infty K_kg^kx^k.
\end{align}
We start with the equation
\begin{equation}
-g^2\frac{\partial^2\Psi_L}{\partial x^2}-V(r)\Psi_L+\big[\omega-qK(r)\big]^2\Psi_L=0.
\label{eq:star}
\end{equation}
Notice that if we let $g=\E^{\iu \cpi/4}$, this reduces to the equation we wish to solve.

As we shall see in the following subsection, it is useful to write $\Psi_L(x)=Y(x)\E^{\beta x^2/2}$ where $\beta$ is a constant to be determined (which depends on $V_j$ and $K_j$, which are known). This will eventually allow us to work with finite order polynomials. We have:
\begin{align}
\Psi_L'(x)&=\Big(Y'(x)+\beta x Y(x)\Big)\E^{\beta x^2/2}, \nonumber \\
\Psi_L''(x)&=\Big(Y''(x)+2\beta x Y'(x)+\beta Y(x)+\beta^2x^2Y(x)\Big)\E^{\beta x^2/2}.
\end{align}
Thus, \eqref{eq:star} becomes
\begin{equation}
-Y''(x)-2\beta x Y'(x)-\beta Y(x)-\beta^2x^2Y(x)+\frac{V_0+V_1gx+\big[\omega-qK(r)\big]^2}{g^2}Y(x)+\sum_{k=2}^\infty V_kg^{k-2}x^kY(x)=0.
\end{equation}
We assign the formal power expansions:
\begin{equation}
Y(x)=\sum_{l=0}^\infty Y_l(x)g^l, \quad \omega=\sum_{n=0}^\infty\omega_ng^n.
\end{equation}
Thus,
\begin{align}
\big[\omega-qK(r)\big]^2 &=\sum_{n=0}^\infty\sum_{m=0}^\infty\big(\omega_n+K_nx^n\big)\big(\omega_m+K_mx^m\big)g^{n+m} \nonumber \\
&=\sum_{l=0}^\infty\sum_{m=0}^l\big(\omega_{l-m}+K_{l-m}x^{l-m}\big)\big(\omega_m+K_mx^m\big)g^l, \quad \text{and hence}
\end{align}
\begin{equation}
\big[\omega-qK(r)\big]^2=\big(\omega_0+K_0\big)^2+2\big(\omega_0+K_0\big)\big(\omega_1+K_1x\big)g+\sum_{l=2}^\infty\sum_{m=0}^l\big(\omega_{l-m}+K_{l-m}x^{l-m}\big)\big(\omega_m+K_mx^m\big)g^l.
\end{equation}
By matching powers of $g$, we require that
\begin{equation}
\big(\omega_0+K_0\big)^2+2\big(\omega_0+K_0\big)\big(\omega_1+K_1x\big)g+V_0+V_1gx=0.
\end{equation}
This implies that $\omega_0=-K_0+s_0\sqrt{-V_0}$, $\omega_1=0$, and $2s_0\sqrt{-V_0}K_1+V_1=0$, where $s_0=\pm1$.

The last condition singles out some points $r_0$ for which, when expanding around $r_0$, we find 
\begin{equation}
2s_0qK'(r_0)\sqrt{V(r_0)}+V'(r_0)=0.
\label{eq:r0extrema}
\end{equation}
In the following, we assume there exists such a point for which the above equation holds. When $qK(r)=0$, as in the uncharged scalar case or a charged scalar in a Schwarzschild black hole background, $r_0$ corresponds to the extremum of the potential, where $V'(r_0)$ vanishes.

Now, assuming the above conditions hold, \eqref{eq:star} becomes:
\begin{align}
-Y''(x)-2\beta x Y'(x)-\beta Y(x)-\beta^2x^2Y(x) +\sum_{k=2}^\infty V_kg^{k-2}x^kY(x) & \nonumber \\ 
+\sum_{l=2}^\infty\sum_{m=0}^l\big(\omega_{l-m}+K_{l-m}x^{l-m}\big)\big(\omega_m+K_mx^m\big)g^{l-2}Y(x)
 &=0.
\end{align}
Replacing $Y$ in the above, we find
\begin{align}
\sum_{l=0}^\infty\bigg\{-Y_l''(x)-2\beta xY_l'(x)-\beta Y_l(x)-\beta^2x^2 Y_l(x)+\sum_{k=0}^lV_{k+2}x^{k+2}Y_{l-k}(x)\bigg\}g^l &\nonumber \\
+\sum_{j=0}^\infty\sum_{n=0}^\infty\sum_{m=0}^{n+2}\big(\omega_{n+2-m}+K_{n+2-m}x^{n+2-m}\big)\big(\omega_m+K_mx^m\big)g^{j+n}Y_j(x) &=0.
\end{align}
Therefore,
\begin{align}
\sum_{l=0}^\infty\bigg\{-Y_l''(x)-2\beta xY_l'(x)-\beta Y_l(x)-\beta^2x^2 Y_l(x)+\sum_{k=0}^lV_{k+2}x^{k+2}Y_{l-k}(x) & \nonumber \\
+\sum_{n=0}^l\sum_{m=0}^{n+2}\big(\omega_{n+2-m}+K_{n+2-m}x^{n+2-m}\big)\big(\omega_m+K_mx^m\big)Y_{l-n}(x)\bigg\}g^l &=0,
\end{align}
which implies that 
\begin{align}
-Y_l''(x)-2\beta xY_l'(x)-\beta Y_l(x)-\beta^2x^2 Y_l(x)+\sum_{k=0}^lV_{k+2}x^{k+2}Y_{l-k}(x) & \nonumber \\
+\sum_{n=0}^l\sum_{m=0}^{n+2}\big(\omega_{n+2-m}+K_{n+2-m}x^{n+2-m}\big)\big(\omega_m+K_mx^m\big)Y_{l-n}(x) &=0,
\label{eq:star2}
\end{align}
for all non-negative integers $l$. In particular, for $l=0$ we have,
\begin{align}
-Y_0''(x)-2\beta xY_0'(x)-\beta Y_0(x)-\beta^2x^2 Y_0(x)+V_2x^2Y_0(x) &\nonumber \\
+\sum_{m=0}^2\big(\omega_{2-m}+K_{2-m}x^{2-m}\big)\big(\omega_m+K_mx^m\big)Y_0(x) &=0,
\end{align}
which, when written out explicitly is
\begin{align}
-Y_0''(x)-2\beta xY_0'(x)-\beta Y_0(x)-\beta^2x^2 Y_0(x)+V_2x^2Y_0(x) & \nonumber \\
+2s_0\sqrt{-V_0}\omega_2Y_0(x)+2s_0\sqrt{-V_0}K_2x^2Y_0(x)+K_1^2x^2Y_0(x) &=0.
\end{align}
Thus,
\begin{equation}
-Y_0''(x)-2\beta xY_0'(x)+\Big(2s_0\sqrt{-V_0}\omega_2-\beta \Big)Y_0(x)+\Big(-\beta^2+V_2+2s_0\sqrt{-V_0}K_2+K_1^2\Big)x^2Y_0(x)=0.
\end{equation}
For the $x^2Y_0(x)$ term to be 0, we require
\begin{equation}
\beta=\pm\sqrt{V_2+K_1^2+2s_0\sqrt{-V_0}K_2}.
\end{equation}
Then, \eqref{eq:star2} for $l=0$ is
\begin{equation}
-Y_0''(x)-2\beta xY_0'(x)=2\Big(\beta/2-s_0\sqrt{-V_0}\omega_2\Big)Y_0(x).
\end{equation}
Comparing with equation 2.10 in \cite{Sulejmanpasic:2016fwr}, we have
\begin{equation}
\omega\rightarrow-\beta \quad \text{and} \quad \epsilon_0\rightarrow-s_0\sqrt{-V_0}\omega_2,
\end{equation}
and so, we find that $Y_0(x)\propto H_\nu(\sqrt{-\beta}x)$ and
\begin{equation}
-s_0\sqrt{-V_0}\omega_2=-\beta\Big(\nu+\frac{1}{2}\Big)\Rightarrow\omega_2=\Big(\nu+\frac{1}{2}\Big)\frac{\beta}{s_0\sqrt{-V_0}}.
\end{equation}
Indeed, letting $t=\sqrt{-\beta}x$ and $Y_0(x)=Z(t(x))$, we find
\begin{align}
Y_0'(x) &=\sqrt{-\beta}Z'(t(x)),  \nonumber \\
Y_0''(x) &=-\beta Z''(t(x)), \nonumber \\
-Y_0''(x)-2\beta xY_0'(x)+\Big(2s_0\sqrt{-V_0}\omega_2-\beta\Big)Y_0(x) &=\beta Z''(t(x))-2\beta\sqrt{-\beta}xZ'(t(x))\nonumber \\ & \quad+\Big(2s_0\sqrt{-V_0}\omega_2-\beta\Big)Z(t(x)).
\end{align}
Thus, \eqref{eq:star2} for $l=0$ becomes
\begin{equation}
Z''(t)-2tZ'(t)=-2\Big(\frac{s_0\sqrt{-V_0}\omega_2}{\beta}-\frac{1}{2}\Big)Z(t),
\end{equation}
which is the Hermite equation, with polynomially bounded solutions
\begin{equation}
Z(t)\propto H_\nu(t), \quad \text{and} \quad \frac{s_0\sqrt{-V_0}\omega_2}{\beta}-\frac{1}{2}=\nu,
\end{equation}
where $\nu$ is a non-negative integer. In the next section, as we develop the recurrence relation for finding $\omega_n$ for $n>2$, we will choose the convention that the proportionality constant is $\big(2\sqrt{-\beta}\big)^{-\nu}$.

In conclusion, we have
\begin{equation}
\omega_2=\Big(\nu+\frac{1}{2}\Big)\frac{\beta}{s_0\sqrt{-V_0}} \quad \text{and} \quad Y_0(x)=\big(2\sqrt{-\beta}\big)^{-\nu}H_\nu\big(\sqrt{-\beta}x\big),
\end{equation}
where $\nu$ is a non-negative integer and the Hermite polynomials are given by 
\begin{equation}
H_0(u)=1, \quad H_{n+1}(u)=2uH_n(u)-H_n'(u).
\end{equation}

As in \cite{Bender:1990pd,Sulejmanpasic:2016fwr}, we choose $\beta < 0$, so that $\Psi_L$  behaves as a decaying Gaussian function and $Y_0$ is real. We also take $s_0 = 1$. We find that choosing the opposite sign, $s_0 = -1$, corresponds to altering the sign convention for $q$. More specifically, we find that:
\begin{align}
\omega(r_0, s_0, \beta, q) &= \omega^*(r_0^*,s_0,-\beta,q), \\
\omega(r_0, s_0, \beta, q) &= -\omega(r_0, -s_0, \beta, -q), 
\label{eq:signchoices}
\end{align}
where $^*$ denotes complex conjugation.

We recall that \eqref{eq:star2} for $l\geq0$ reads
\begin{align}
-Y_l''(x)-2\beta xY_l'(x)-\beta Y_l(x)-\beta^2x^2 Y_l(x)+\sum_{k=0}^lV_{k+2}x^{k+2}Y_{l-k}(x) & \nonumber \\
+\sum_{n=0}^l\sum_{m=0}^{n+2}\big(\omega_{n+2-m}+K_{n+2-m}x^{n+2-m}\big)\big(\omega_m+K_mx^m\big)Y_{l-n}(x) &=0.
\end{align}
The equation can then be rewritten:
\begin{align}
-Y_l''(x)-2\beta xY_l'(x)-\beta Y_l(x)+2s_0\sqrt{-V_0}\omega_2Y_l(x)+\sum_{k=1}^lV_{k+2}x^{k+2}Y_{l-k}(x) & \nonumber \\
+\sum_{n=1}^l\sum_{m=0}^{n+2}\big(\omega_{n+2-m}+K_{n+2-m}x^{n+2-m}\big)\big(\omega_m+K_mx^m\big)Y_{l-n}(x) &=0.
\end{align}
Or, replacing $\omega_2$ in the above we find:
\begin{align}
-Y_l''(x)-2\beta xY_l'(x)+2\nu\beta Y_l(x)+\sum_{k=1}^lV_{k+2}x^{k+2}Y_{l-k}(x) & \nonumber \\
+\sum_{n=1}^l\sum_{m=0}^{n+2}\big(\omega_{n+2-m}+K_{n+2-m}x^{n+2-m}\big)\big(\omega_m+K_mx^m\big)Y_{l-n}(x) &=0.
\end{align}
For all integers $l$, let
\begin{equation}
Y_l(x)=\sum_{k=0}^{\infty}A_l^kx^k,
\end{equation}
where $A_l^k$ are constants to be determined. We set $A_{-n}^k=0$ for all $n\geq1$ and $A_l^{-m}=0$ for all $m\geq1$. Since we have $Y_0(x)=\big(2\sqrt{-\beta}\big)^{-\nu}H_\nu\big(\sqrt{-\beta}x\big)$, $Y_0$ is an order $\nu$ polynomial. Thus, $A_0^k=0$ for $k>\nu$.

We can also show inductively that if we assume that $Y_l$ is polynomially bounded, then $A_l^k=0$ for all $k>\nu+3l$: By the induction hypothesis, the last two terms can only go up to order $\nu+3(l-1)+3=\nu+3l$. Suppose there exists a finite highest order $M>\nu+3l$ for which $A_l^M\not=0$. Then, we find that $A_l^M$ must satisfy
\begin{equation}
2\beta A_l^M\big(\nu-M\big)=0,
\end{equation}
contradicting the assumption that $A_l^M\not=0$ (clearly $\beta$ cannot be 0 and we required $M>\nu+3l$). Thus, there cannot be a finite highest order $M>\nu+3l$ for which $A_l^M\not=0$, so if $Y_l$ is polynomially bounded, then $A_l^k=0$ for all $k>\nu+3l$. In the following we assume this is the case. Then, \eqref{eq:star2} becomes:
\begin{align}
\sum_{k=0}^\infty \bigg\{-(k+1)(k+2)A_l^{k+2}-2k\beta A_l^k+2\nu\beta A_l^k\bigg\}x^k+\sum_{n=1}^l\sum_{m=0}^\infty V_{n+2}A_{l-n}^mx^{n+m+2} & \nonumber \\
+\sum_{n=1}^l\sum_{m=0}^{n+2}\big(\omega_{n+2-m}+K_{n+2-m}x^{n+2-m}\big)\big(\omega_m+K_mx^m\big)Y_{l-n}(x) &=0.
\end{align}
Then,
\begin{align}
\sum_{k=0}^\infty \bigg\{-(k+1)(k+2)A_l^{k+2}+2(\nu-k)\beta A_l^k+\sum_{n=1}^l V_{n+2}A_{l-n}^{k-n-2}\bigg\}x^k & \nonumber \\
+\sum_{n=1}^l\sum_{m=0}^{n+2}\sum_{j=0}^\infty \big(\omega_{n+2-m}\omega_m+2\omega_{n+2-m}K_mx^m+K_{n+2-m}K_mx^{n+2}\big)A_{l-n}^jx^j  & =0. \nonumber \\
\sum_{k=0}^\infty \bigg\{-(k+1)(k+2)A_l^{k+2}+2(\nu-k)\beta A_l^k+\sum_{n=1}^l V_{n+2}A_{l-n}^{k-n-2}+\sum_{n=1}^l\sum_{m=0}^{n+2}\omega_{n+2-m}\omega_mA_{l-n}^k\bigg\}x^k &  \nonumber \\
+\sum_{j=0}^\infty\sum_{n=1}^l\sum_{m=0}^{n+2}\big(2\omega_{n+2-m}K_mA_{l-n}^jx^{m+j}+K_{n+2-m}K_mA_{l-n}^jx^{n+j+2}\big) & =0.
\end{align}
Thus, we find that
\begin{align}
\sum_{k=0}^\infty \bigg\{-(k+1)(k+2)A_l^{k+2}+2(\nu-k)\beta A_l^k+\sum_{n=1}^l V_{n+2}A_{l-n}^{k-n-2}+\sum_{n=1}^l\sum_{m=0}^{n+2}\omega_{n+2-m}\omega_mA_{l-n}^k &\nonumber \\
+\sum_{n=1}^l\sum_{m=0}^{n+2}2\omega_{n+2-m}K_mA_{l-n}^{k-m}+\sum_{n=1}^l\sum_{m=0}^{n+2}K_{n+2-m}K_mA_{l-n}^{k-n-2}\bigg\}x^k &=0,
\end{align}
which implies that for all $k\geq0$, the following equation holds:
\begin{align}
-(k+1)(k+2)A_l^{k+2}+2(\nu-k)\beta A_l^k+\sum_{n=1}^l V_{n+2}A_{l-n}^{k-n-2}+\sum_{n=1}^l\sum_{m=0}^{n+2}\omega_{n+2-m}\omega_mA_{l-n}^k & \nonumber \\
+\sum_{n=1}^l\sum_{m=0}^{n+2}2\omega_{n+2-m}K_mA_{l-n}^{k-m}+\sum_{n=1}^l\sum_{m=0}^{n+2}K_{n+2-m}K_mA_{l-n}^{k-n-2} &=0.
\label{eq:star3}
\end{align}

\subsection{Finding the Coefficients of the Power Series Expansion}

We start with \eqref{eq:star3}, where $A_l^k=0$ for all $l<0$, $A_l^k=0$ for all $k<0$ or $k>\nu+3l$. To illustrate these constraints explicitly, we can rewrite \eqref{eq:star3} as
\begin{align}
-(k+1)(k+2)A_l^{k+2}+2(\nu-k)\beta A_l^k+\sum_{n=1}^{N_1} \Big(V_{n+2}+\sum_{m=0}^{n+2}K_{n+2-m}K_m\Big)A_{l-n}^{k-n-2} & \nonumber \\
+\sum_{n=1}^{N_2}\sum_{m=0}^{n+2}\omega_{n+2-m}\omega_mA_{l-n}^k+\sum_{n=1}^l\sum_{m=N_3}^{N_4}2\omega_{n+2-m}K_mA_{l-n}^{k-m}&=0,
\end{align}
where
\begin{align}
N_1 &=\min\Big(k-2,l,\frac{\nu+3l+2-k}{2}\Big), \nonumber \\
N_2 &=\min\Big(l,l+\frac{\nu-k}{3}\Big), \nonumber \\
N_3 &=\max\big(0,k-\nu-3l+3n\big), \nonumber \\
N_4 &=\min\big(k,n+2\big).
\end{align}
If we let $k=\nu$, the equation becomes
\begin{align}
-(\nu+1)(\nu+2)A_l^{\nu+2}+\sum_{n=1}^{\min(\nu-2,l)} \Big(V_{n+2}+\sum_{m=0}^{n+2}K_{n+2-m}K_m\Big)A_{l-n}^{\nu-n-2} & \nonumber \\
+\sum_{n=1}^l\sum_{m=0}^{n+2}\omega_{n+2-m}\omega_mA_{l-n}^\nu+\sum_{n=1}^l\sum_{m=0}^{\min(\nu,n+2)}2\omega_{n+2-m}K_mA_{l-n}^{\nu-m} &=0,
\end{align}
which can be rewritten as
\begin{align}
-(\nu+1)(\nu+2)A_l^{\nu+2}+\sum_{n=1}^{\min(\nu-2,l)} \Big(V_{n+2}+\sum_{m=0}^{n+2}K_{n+2-m}K_m\Big)A_{l-n}^{\nu-n-2} &\nonumber \\
+\sum_{n=1}^{l-1}\sum_{m=0}^{n+2}\omega_{n+2-m}\omega_mA_{l-n}^\nu
+2\omega_{l+2}\big(\omega_0+K_0\big)A_0^\nu +\sum_{m=1}^{l+1}\omega_{l+2-m}\omega_mA_0^\nu & \nonumber \\
+\sum_{n=1}^{l-1}\sum_{m=0}^{\min(\nu,n+2)}2\omega_{n+2-m}K_mA_{l-n}^{\nu-m}+\sum_{m=1}^{\min(\nu,l+2)}2\omega_{l+2-m}K_mA_0^{\nu-m} &=0.
\end{align}
This shows that knowing all $A_j^k$ for $0\leq j<l$, $A_l^{\nu+2}$, and all $\omega_j$ for $0\leq j\leq l+1$ leads to a formula for $\omega_{l+2}$:
\begin{align}
\omega_{l+2}=-\frac{1}{2s_0\sqrt{-V_0}A_0^\nu}\Bigg\{ & -(\nu+1)(\nu+2)A_l^{\nu+2}+\sum_{n=1}^{\min(\nu-2,l)} \Big(V_{n+2}+\sum_{m=0}^{n+2}K_{n+2-m}K_m\Big)A_{l-n}^{\nu-n-2} \nonumber \\
& +\sum_{n=1}^{l-1}\sum_{m=0}^{n+2}\omega_{n+2-m}\omega_mA_{l-n}^\nu+\sum_{m=1}^{l+1}\omega_{l+2-m}\omega_mA_0^\nu \nonumber \\
& +\sum_{n=1}^{l-1}\sum_{m=0}^{\min(\nu,n+2)}2\omega_{n+2-m}K_mA_{l-n}^{\nu-m}+\sum_{m=1}^{\min(\nu,l+2)}2\omega_{l+2-m}K_mA_0^{\nu-m}\Bigg\}.
\label{eq:hash}
\end{align}

We additionally set $A_0^\nu=1$ and $A_l^\nu=0$ for $l\geq1$ as a normalization condition. This sets the proportionality constant for $Y_0$, $Y_0(x)=\big(2\sqrt{-\beta}\big)^{-\nu}H_\nu\big(\sqrt{-\beta}x\big)$.

Now that we know how to find the components of $\omega$, we need to find the $A_l^k$ for $l\geq1$. To do this, we rewrite \eqref{eq:star3} for $k\not=\nu$ as 
\begin{align}
A_l^k =\frac{1}{2(k-\nu)\beta}\Bigg\{& -(k+1)(k+2)A_l^{k+2}+\sum_{n=1}^{N_1} \left(V_{n+2}+\sum_{m=0}^{n+2}K_{n+2-m}K_m\right)A_{l-n}^{k-n-2} \nonumber \\
&+ \sum_{n=1}^{N_2}\sum_{m=0}^{n+2}\omega_{n+2-m}\omega_mA_{l-n}^k+ 
\sum_{n=1}^l\sum_{m=N_3}^{N_4}2\omega_{n+2-m}K_mA_{l-n}^{k-m}\Bigg\}.
\label{eq:hash2}
\end{align}
For $k>\nu$, we simply start from $k=\nu+3l$ and go down to $k=\nu+1$ since we know that $A_l^{k+2}=0$ for the first few values for which $k>\nu+3l-2$. An important observation is that in the above formula for $k>\nu$, $\omega_{l+2}$ does not actually appear since $\omega_{l+2}$ multiplies $A_0^k$ which is $0$ for $k>\nu$. Once one has found all $A_l^k$ for $k>\nu$, one can find $\omega_{l+2}$ from \eqref{eq:hash} and then use \eqref{eq:hash2} from $k=\nu-1$ down to $k=0$ to find $A_l^k$ for $0\leq k<\nu$.

After we have found these recurrence relations, we can also prove an interesting simplification by induction: $\omega_l=0$ for $l$ odd and $A_l^k=0$ for $k-l-\nu$ odd. This can be seen as follows. First, note that either all the coefficients of the odd powers of $x$ or all the coefficients of the even powers of $x$ in a Hermite polynomial are 0. Since $A_0^\nu\not=0$, this means that $A_0^k=0$ for $k-\nu$ odd. Also, we saw that $\omega_1=0$, proving the base case. Now let $l$ be a positive integer. We assume that for all $j\leq l$ if $j$ is odd then $\omega_j=0$, and for all $j<l$ and all $k$ if $k-j-\nu$ is odd then $A_j^k=0$ as the induction hypothesis and want to show that $\omega_{l+2}=0$ if $l$ is odd and that $A_l^k=0$ if $k-l-\nu$ is odd. This is straightforward to see directly from the above recurrence relations. We begin at $k=\nu+3l$ and work our way down to $k=\nu+1$. We assume that $k-l-\nu$ is odd. For $k=\nu+3l$, $A_l^{k+2}=0$ trivially so the first term vanishes. For $\nu+1\leq k<\nu+3l$ we would have already proven that $A_l^{k+2}=0$ since $k-l-\nu$ odd implies that $(k+2)-l-\nu$ odd. Thus, the first term is always 0. Also, $k-l-\nu$ odd implies that $(k-n-2)-(l-n)-\nu$ odd so the second term is always 0. Also for $\omega_{n+2-m}\omega_mA_{l-n}^k$ to be nonzero, we would require $n+2-m$ even, $m$ even, $k-(l-n)-\nu$ even, which implies $n$ even and so $k-l-\nu$ even, which is false. Thus, the third term is also 0. For $\omega_{n+2-m}A_{l-n}^{k-m}$ to be nonzero, we would require $n+2-m$ even and $(k-m)-(l-n)-\nu$ even which would imply $k-l-\nu$ even, a contradiction. Thus, the fourth term is also 0. Thus, we have shown the claim for $k>\nu$. We now need to show that $\omega_{l+2}=0$ if $l$ is odd. $(\nu+2)-l-\nu$ is odd so $A_l^{\nu+2}=0$ as we have just proven, so the first term is 0. $(\nu-n-2)-(l-n)-\nu$ is odd since $l$ is odd so $A_{l-n}^{\nu-n-2}=0$, so the second term is also 0. For $\omega_{n+2-m}\omega_mA_{l-n}^\nu$ to be nonzero we would require $n+2-m$ even, $m$ even, $\nu-(l-n)-\nu$ even, so $l$ even, contradiction. Thus, the third and fourth terms are 0. For $\omega_{n+2-m}A_{l-n}^{\nu-m}$ to be nonzero, we would require $n+2-m$ even and $(\nu-m)-(l-n)-\nu$ even, so $l$ even, contradiction. Thus, the fifth and sixth terms are 0. Thus $\omega_{l+2}=0$. Now it is analogous to show that $k-l-\nu$ odd implies $A_l^k=0$ for $k<\nu$. This concludes the induction step and the proof. 

\subsection{Comments on boundary conditions}
\label{subsec:bcs}

Our algorithm provides an approximate solution to \eqref{eq:star}, by analogy to the method of \cite{Hatsuda:2019eoj}. To argue that these solutions evaluated at $g = \E^{\iu \cpi/4}$ are the quasinormal modes, rather than some other set of solutions, we should consider the boundary conditions that they obey. While we do not have a rigorous proof that the method we propose converges to the correct solutions, the argument that we expect them to do so in general closely follows that given below equation (2.3) of \cite{Hatsuda:2019eoj}. Although we have derived all expressions as a series in $g$, in the end the odd terms vanish and the expansion is in $g^2$, precisely analogous (and in the $q = 0$ case, identical) to Hatsuda's expansion in $\hbar$. We will now give a more detailed explanation.

To study boundary conditions for solutions outside the black hole, we are interested in the behavior of the functions $F(r)$ and $K(r)$ as $r \to \infty$ and $r \to r_+$. We will not solve the equation in general, but specifically for the case of asymptotically flat Reissner--Nordstr{\"o}m black holes \eqref{eq:RNsolution}. In this case, the function $F(r)$ has the asymptotic limits
\begin{align}
r \to \infty: &\qquad F(r) \to 1 - \left(\frac{r_S}{r}\right)^{D-3} + \cdots, \nonumber \\
r \to r_+: &\qquad F(r) \to 0 + (D-3) \left(1 - \left(\frac{r_-}{r_+}\right)^{D-3}\right) \left(\frac{r}{r_+} - 1\right) + \cdots.
\end{align}
The qualitative behavior is not specific to Reissner--Nordstr{\"o}m black holes: the $r \to \infty$ behavior is the usual expectation for an asymptotically flat metric, where the coefficient of the subleading term is proportional to the ADM mass $M$ (see \eqref{eq:RNMQ}), while the vanishing of $F$ as $r \to r_+$ is characteristic of the presence of a horizon. Meanwhile, the function $K(r) = -A_t(r)$ in \eqref{eq:RNvectorpotential} also approaches constant values in the two limits:
\begin{align}
r \to \infty: &\qquad K(r) \to K_\infty +  \frac{Q}{(D-3) \Omega_{D-2}} \frac{1}{r^{D-3}}, \nonumber \\
r \to r_+: &\qquad K(r) \to K_+ - \frac{Q}{\Omega_{D-2}} \frac{1}{r_+^{D-3}} \left(\frac{r}{r_+} - 1\right) + \cdots, 
\end{align}
where only the {\em difference} of $K_\infty$ and $K_+$ is physical:
\begin{equation}
K_\infty - K_+ = - \frac{Q}{(D-3)\Omega_{D-2}} \frac{1}{r_+^{D-3}}.
\end{equation}
The constant $K_\infty$ is a gauge choice, a point to which we will return at the beginning of \S\ref{sec:validateDMs}. 

In both asymptotic regions, the leading behavior of the solution is determined by solving a differential equation of the form
\begin{equation}
\left[-g^2 \frac{{\rm d}^2}{{\rm d}x^2} + \left(\omega - q K\right)^2 - \lambda^2\right] \Psi_L(x) = 0,
\end{equation}
where $K$ is either $K_\infty$ or $K_+$ depending on the region under consideration, and $\lambda$ is $\mu$ for $r \to \infty$ and $0$ for $r \to r_+$. In the $y \to -\infty$ ($r \to r_+$) region, we have $x \sim y/g \propto \log(r - r_+)$, so the corrections to this equation are {\em exponentially} small. In the $y \to \infty$ region, $x$ and $r$ are linearly related, and the corrections are suppressed by powers of $x$. This implies that the asymptotic dependence of the solution has the form:
\begin{align}
y \to -\infty: &\quad \Psi_L(x) \sim \exp\left[\pm \frac{x}{g} (\omega - q K_+)\right] \sim \exp\left[\pm \frac{y}{g^2} (\omega - qK_+) \right], \nonumber \\
y \to +\infty: &\quad \Psi_L(x) \sim A(x) \exp\left[\pm \frac{x}{g} \sqrt{(\omega - qK_\infty)^2 - \mu^2}\right] \sim A(y/g) \exp\left[\pm \frac{y}{g^2} \sqrt{(\omega - qK_\infty)^2 - \mu^2}\right], 
  \label{eq:solutionasymptotics}
\end{align}
where $A(y/g)$ varies more slowly than the exponential factor (it arises from the subleading power-law terms as $x \to \infty$, and scales asymptotically as an $\omega$-dependent power of $y$).

The solutions that we have constructed, for $g$ real and positive, take the form of a decaying Gaussian $\exp(\beta x^2/2)$ (with $\beta < 0$) multiplied by more slowly-varying functions. Thus, while they may not have the same asymptotic form as \eqref{eq:solutionasymptotics}, to the extent that they are approximating solutions of the same equation, they will only provide a good approximation to the solutions that are exponentially decaying in both asymptotic regions, i.e., the choice with a $+$ sign for $y \to -\infty$ and a $-$ sign for $y \to +\infty$ in \eqref{eq:solutionasymptotics} when picking the positive branch of the square root. Then, we analytically continue from real, positive $g$ to $g = \E^{\iu \cpi/4}$, assuming (as discussed in \cite{Hatsuda:2019eoj}) that there is no Stokes phenomenon spoiling this continuation. Then $g^2 \mapsto \iu$ in the equations, and we obtain (for the analytically continued result)
\begin{align}
y \to -\infty: &\qquad \Psi_L(y) \sim  \exp\left[- \iu y\, (\omega|_{g^2 \mapsto \iu} - qK_+)\right], \nonumber \\
y \to +\infty: &\qquad \Psi_L(y) \sim A(\E^{-\iu \cpi/4} y)\exp\left[\iu y \sqrt{(\omega|_{g^2 \mapsto \iu} - qK_\infty)^2 - \mu^2}\right].
  \label{eq:finalsolutionasymptotics}
\end{align}
We use the notation $\omega|_{g^2 \mapsto \iu}$ to emphasize that we find solutions for $\omega$ as a series in $g$, which we then evaluate at the analytically continued choice $g = \exp(\iu \cpi/4)$. The branch of the square root is that determined by starting with the positive root when $g$ is real and positive, and analytically continuing. The behavior we find in \eqref{eq:finalsolutionasymptotics} is the expected asymptotic behavior of a quasinormal mode solution, which we reviewed in the introduction below \eqref{eq:omRomI}. Hence, we expect that the algorithm we have described will obtain the correct solutions (although we do not have a rigorous proof). We will validate this expectation with numerical results. The argument that we have given applies specifically to solutions that, for real and positive $g$, have the form of exponentially decaying Gaussians {\em outside} the black hole horizon, so that we are solving the equation in the region $r_+ < r < \infty$. Below, we will also present some results obtained by expanding around a point $r_0$ satisfying $r_0 < r_+$. In that case, we have only empirical findings, not a clear understanding of the boundary conditions that the modes obey when extrapolated to $r > r_+$.

\subsection{Summary of the Method}
\label{subsec:summary}

We want to numerically solve for the modes $\omega$ of the equation
\begin{equation}
\Big\{\big[\omega-qK(r)\big]^2+\partial_y^2-V(r)\Big\}\Psi_L(r)=0,
\end{equation}
where
{\small
\begin{equation}
V(r)=F(r)\Bigg\{\mu^2+\frac{l_{D-2}\big(l_{D-2}+D-3\big)}{H(r)}+\frac{D-2}{16G(r)}\Big[(D-6)\frac{H'(r)^2}{H(r)^2}+4\frac{H''(r)}{H(r)}+2\frac{H'(r)}{H(r)}\Big(\frac{F'(r)}{F(r)}-\frac{G'(r)}{G(r)}\Big)\Big]\Bigg\},
\end{equation}
}
and $r=r(y)$, with $r'(y)=\sqrt{\frac{F(r)}{G(r)}}$.

Let $s_0=\pm1$, and $r_0=r(y_0)$ be a point for which
\begin{equation}
2s_0qK'(r_0)\sqrt{V(r_0)}+V'(r_0)=0.
\end{equation}
Let $y-y_0=gx$, and
\begin{equation}
-V(r)=\sum_{k=0}^\infty V_kg^kx^k,\quad -qK(r)=\sum_{k=0}^\infty K_kg^kx^k.
\end{equation}
Let $\Psi_L(x)=Y(x)\E^{\beta x^2/2}$ where $\beta=\pm\sqrt{V_2+K_1^2+2s_0\sqrt{-V_0}K_2}$, and
\begin{equation}
Y(x)=\sum_{l=0}^\infty Y_l(x)g^l,\quad  \omega=\sum_{n=0}^\infty\omega_ng^n.
\end{equation}
where
\begin{equation}
\omega_0=-K_0+s_0\sqrt{-V_0},\quad\omega_1=0,\quad\omega_2=\Big(\nu+\frac{1}{2}\Big)\frac{\beta}{s_0\sqrt{-V_0}},\quad\text{and}\quad Y_0(x)=\big(2\sqrt{-\beta}\big)^{-\nu}H_\nu\big(\sqrt{-\beta}x\big),
\end{equation}
where $\nu$ is a non-negative integer and the Hermite polynomials are given by 
\begin{equation}
H_0(u)=1, \quad H_{n+1}(u)=2uH_n(u)-H_n'(u).
\end{equation}
For all integers $l\geq0$, let
\begin{equation}
Y_l(x)=\sum_{k=0}^{\nu+3l}A_l^kx^k,
\end{equation}
where $A_l^k=0$ if $l<0$ or $k<0$ or $k>\nu+3l$ or $l\geq1$ and $k=\nu$.
Since $Y_0$ is a known polynomial of order $\nu$, we find $A_0^k$ for $0\leq k\leq\nu$. Then, for all $l\geq 1$, we find $A_l^k$ for $0\leq k\leq \nu+3l$ using  the following procedure.

Let
\begin{align}
N_1&=\min\Big(k-2,l,\frac{\nu+3l+2-k}{2}\Big),\nonumber\\
N_2&=\min\Big(l,l+\frac{\nu-k}{3}\Big), \nonumber \\
N_3&=\max\big(0,k-\nu-3l+3n\big), \nonumber \\
N_4&=\min\big(k,n+2\big).
\end{align}
From $k=\nu+3l$ down to $k=\nu+1$, we have
\begin{align}
A_l^k =\frac{1}{2(k-\nu)\beta}\Bigg\{& -(k+1)(k+2)A_l^{k+2}+\sum_{n=1}^{N_1} \left(V_{n+2}+\sum_{m=0}^{n+2}K_{n+2-m}K_m\right)A_{l-n}^{k-n-2} \nonumber \\
&+ \sum_{n=1}^{N_2}\sum_{m=0}^{n+2}\omega_{n+2-m}\omega_mA_{l-n}^k+ 
\sum_{n=1}^l\sum_{m=N_3}^{N_4}2\omega_{n+2-m}K_mA_{l-n}^{k-m}\Bigg\}.
\label{eq:Akl}
\end{align}
After using this relation to find $A_l^k$ for $k$ from $\nu+3l$ down to $\nu+1$, we use
\begin{align}
\omega_{l+2}=-\frac{1}{2s_0\sqrt{-V_0}A_0^\nu}\Bigg\{ & -(\nu+1)(\nu+2)A_l^{\nu+2}+\sum_{n=1}^{\min(\nu-2,l)} \Big(V_{n+2}+\sum_{m=0}^{n+2}K_{n+2-m}K_m\Big)A_{l-n}^{\nu-n-2} \nonumber \\
& +\sum_{n=1}^{l-1}\sum_{m=0}^{n+2}\omega_{n+2-m}\omega_mA_{l-n}^\nu+\sum_{m=1}^{l+1}\omega_{l+2-m}\omega_mA_0^\nu \nonumber \\
& +\sum_{n=1}^{l-1}\sum_{m=0}^{\min(\nu,n+2)}2\omega_{n+2-m}K_mA_{l-n}^{\nu-m}+\sum_{m=1}^{\min(\nu,l+2)}2\omega_{l+2-m}K_mA_0^{\nu-m}\Bigg\}
\end{align}
to find $\omega_{l+2}$, after which we use \eqref{eq:Akl} again to find $A_l^k$ for $k$ from $\nu-1$ down to $0$. We also see from these equations that $\omega_l=0$ for odd $l$ and that $A_l^k=0$ for $k-l-\nu$ odd.
This concludes the procedure and lets us find $\omega_l$ for arbitrarily large $l$. We then implement Hatsuda's technique, with $g^2=\E^{\iu \cpi/2}=\iu$ and find $\omega$.

\section{Validating the algorithm for damped modes}
\label{sec:validateDMs}

In this section, we show that the algorithm discussed in the previous section can compute the frequencies of damped quasinormal modes of charged scalar fields in the 4d Reissner--Nordstr{\"o}m background. These damped modes have $\omI \neq 0$ in the extremal limit. We specified the Reissner--Nordstr{\"o}m solution, including the gauge field strength $F_{tr}$, in \eqref{eq:RNsolution}. This leaves a gauge ambiguity in the choice of the function $K(r) = -A_t$. In what follows, we fix this function to be
\begin{equation}
K(r) = \frac{Q}{(D-3)\Omega_{D-2}} \left(\frac{1}{r^{D-3}}  - \frac{1}{r_+^{D-3}}\right),
\end{equation} 
vanishing at the outer horizon. In some of the literature, the second term in parentheses is dropped, so that $K(r) \to 0$ at $r \to \infty$. This mismatch in conventions corresponds to a constant shift in the real part of the quasinormal mode frequency for a given field, because only the  combination $\omega - q K(r)$ appears in the equations of motion. Denoting the alternative values obtained in the literature by ${\widetilde \omega}$, we have
\begin{equation}
{\widetilde \omega}_\mathrm{R} = \omR + \frac{q Q}{(D-3)\Omega_{D-2} r_+^{D-3}}.
\end{equation}

Damped modes in Reissner--Nordstr{\"o}m backgrounds have previously been studied from a variety of perspectives: with WKB approximations \cite{Konoplya:2002ky,Konoplya:2002wt}, numerical simulations of time-dependent solutions \cite{Xue:2002xa}, and Leaver's method of continued fractions \cite{Konoplya:2007zx, Konoplya:2013rxa, Richartz:2014jla, Chowdhury:2018izv}. To validate the results of calculations obtained with our algorithm, we compare to results we have obtained using Leaver's method (including Nollert's improvement, which estimates the remaining part of the truncated continued fraction \cite{Nollert:1993zz}). We have checked that our implementations of both Leaver's method and our new algorithm are able to reproduce a variety  of numerical results and plots from \cite{Konoplya:2002ky,Konoplya:2002wt, Chowdhury:2018izv},  including the generalized results with a scalar hair parameter from \cite{Chowdhury:2018izv}. 

\begin{figure}[!h]
\centering
\includegraphics[width=\textwidth]{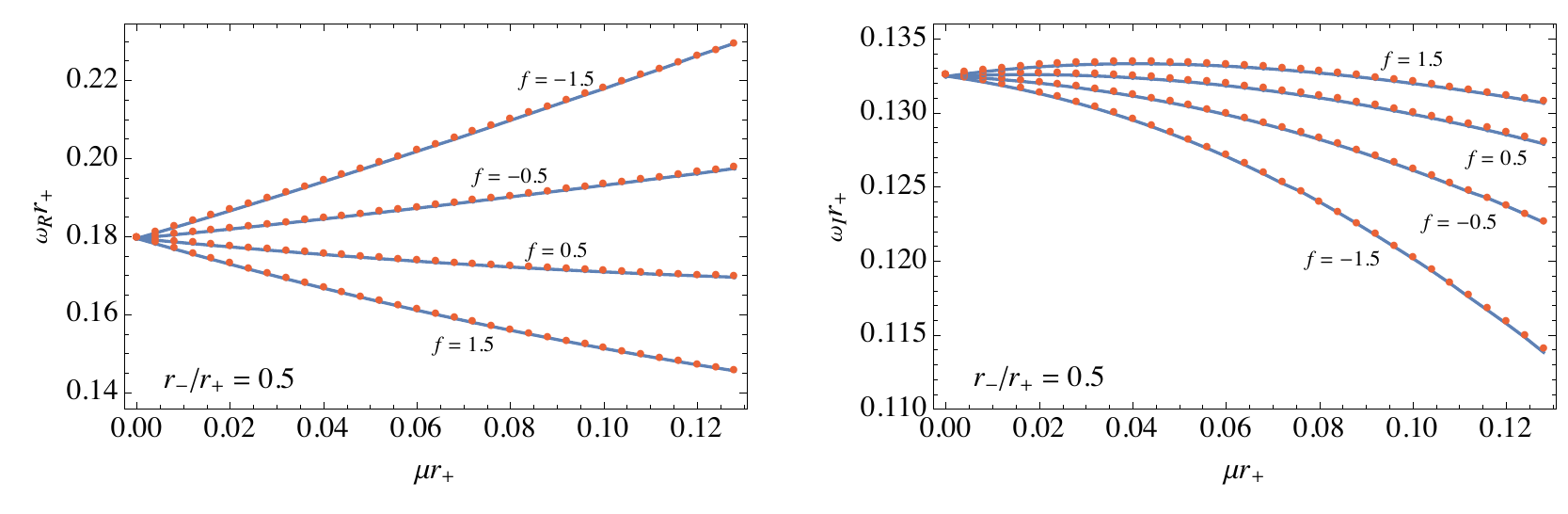}
\caption{Dependence of the real and imaginary parts of the lowest-lying $l = 0$ quasinormal mode frequency of a charged scalar on the scalar mass parameter $\mu$ (horizontal axis) and its charge-to-mass ratio $f$ (as labeled on each curve). We measure  all quantities in units set by the outer horizon radius, $r_+$,  and have fixed $r_-/r_+ = 1/2$. The continuous blue lines interpolate between results obtained with our extension of Hatsuda's algorithm, while the red dots were obtained with Leaver's method. They agree within the precision of the calculation.}
\label{fig:mudependencevaryingf}
\end{figure} 

Here, we will present several comparison plots showing numerical results that we have obtained with the algorithm of \S\ref{sec:algorithm} and with Leaver's method. We restrict our study to the case $D = 4$, and plot modes of $l = 0$. The quasinormal mode problem is characterized by four parameters ($r_-, r_+, q/\kappa, \mu$), but one combination of parameters simply sets the overall scale of the problem, so that the physically meaningful parameters can be viewed as $\mu r_+, r_-/r_+,$ and $q/\kappa$, and the output  as $\omega r_+$. We find it convenient to parametrize the scalar field charge in terms of its fraction of the extremality bound, which we denote
\begin{equation}
f \df  \sqrt{\frac{D-2}{D-3}} \frac{q}{\kappa \mu}.
\end{equation}
A scalar field that satisfies the Weak Gravity Conjecture is one for which $|f| \geq 1$ \cite{ArkaniHamed:2006dz, Heidenreich:2015nta}. As a first illustration of our results, in Fig.~\ref{fig:mudependencevaryingf} we fix $r_-/r_+ = 1/2$ and show how the quasinormal mode frequencies depend on $\mu r_+$ for several discrete choices of $f$. We find good agreement between our method and Leaver's method, within the numerical precision to which we are working.

\begin{figure}[!h]
\centering
\includegraphics[width=\textwidth]{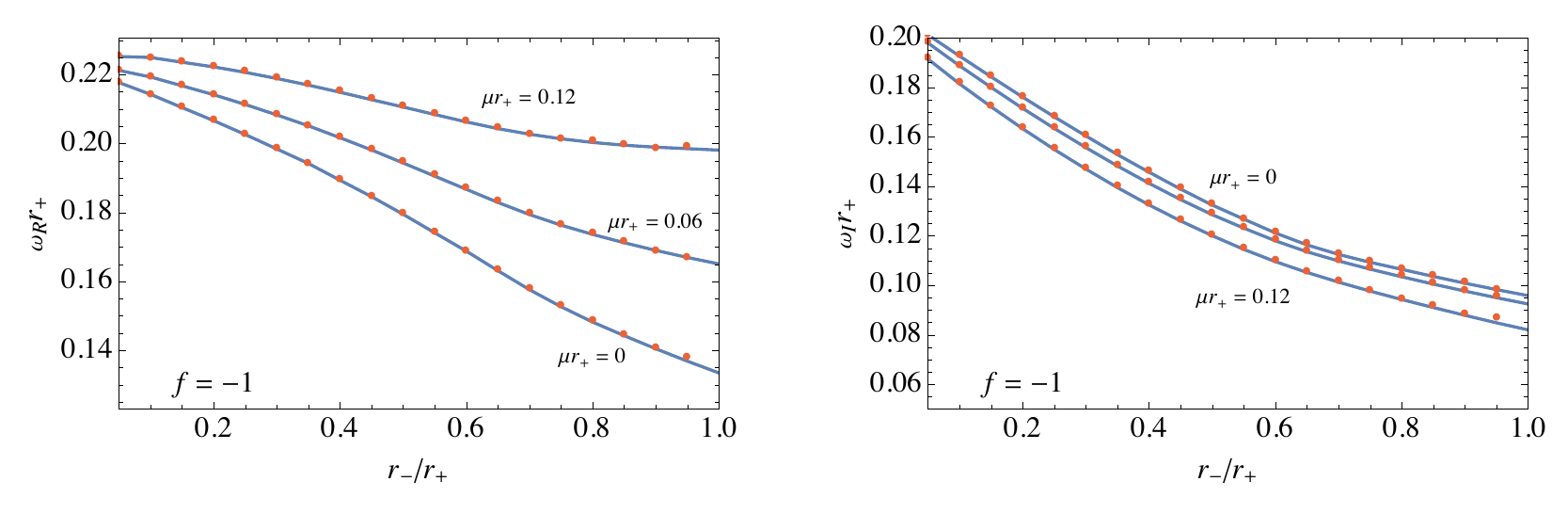}
\caption{Dependence of the real and imaginary parts of the lowest-lying $l = 0$ quasinormal mode frequency of a charged scalar on the ratio of horizon coordinates $r_-/r_+$ and the scalar mass parameter $\mu$ (as labeled on each curve). We have fixed the scalar charge fraction $f = -1$. Blue lines are our algorithm and red dots are Leaver's method. In the right-hand plot, we have suppressed the label $\mu r_+ = 0.06$ on the middle curve for legibility.}
\label{fig:rmdependencevaryingmu}
\end{figure} 

\begin{figure}[!h]
\centering
\includegraphics[width=\textwidth]{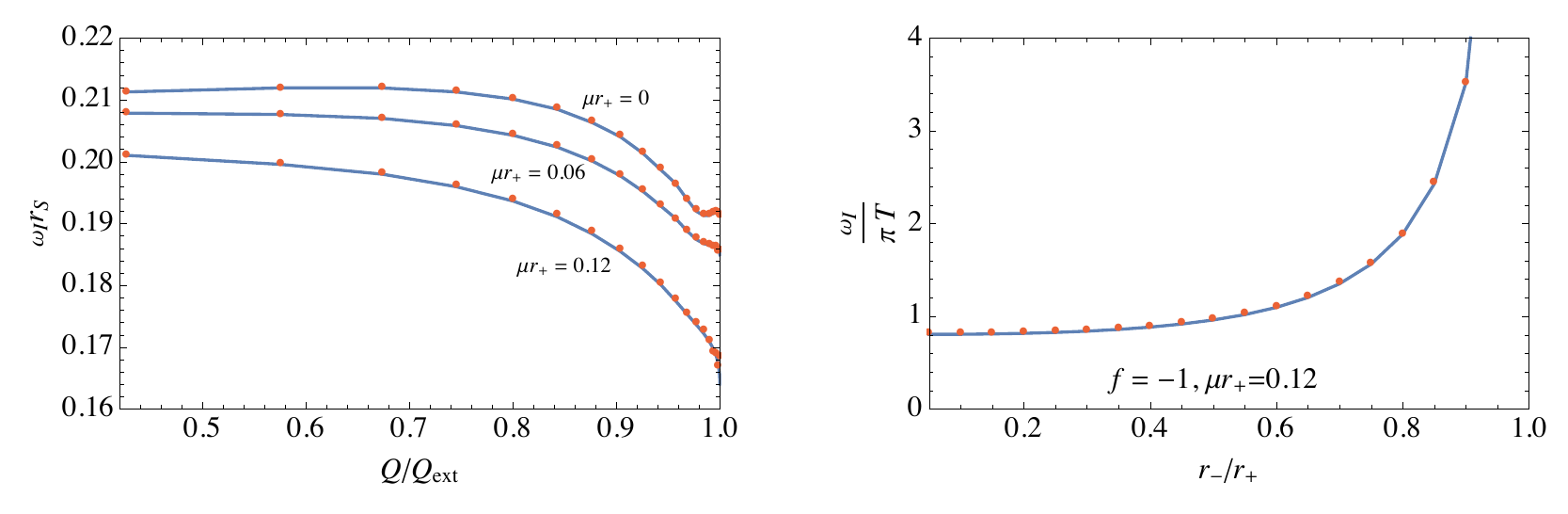}
\caption{Dependence of the imaginary part of the lowest-lying $l = 0$ quasinormal mode frequency of a massless scalar, $\omI$, measured in different units. At left, we show $\omI r_S$ as a function of $Q/Q_\mathrm{ext}$ for the same choices of $\mu r_+$ shown in Fig.~\ref{fig:rmdependencevaryingmu}. At right, we show $\omI/(\cpi T)$ as a function of $r_-/r_+$, selecting a particular $\mu$ since the curves of different $\mu$ are difficult to separate when plotted in this manner.}
\label{fig:omegaIalternative}
\end{figure}

In Fig.~\ref{fig:rmdependencevaryingmu}, we show similar results, this time as a function of the ratio $r_-/r_+$ and for three choices of $\mu r_+$. Again, we find good agreement between our generalization of Hatsuda's method and our implementation of Leaver's method. Notice that our generalization of Hatsuda's method can produce results even at extremality, when $r_- = r_+$, and just as we showed for the gravitational and electromagnetic perturbations in \S\ref{subsec:opportunities}, $\omega_I$ remains finite in this limit. Again, this indicates that the algorithm is finding DMs rather than ZDMs. In Fig.~\ref{fig:omegaIalternative}, we have re-plotted the same information in different ways: first, we show $\omI r_S$ versus the extremality fraction $Q/Q_\mathrm{ext}$; second, we show $\omI/(\cpi T)$ versus $r_-/r_+$. For the example that we have plotted, we see that the damped modes violate the relaxation bound $\omI \leq \cpi T$ \cite{Hod:2006jw} over a substantial part of parameter space, so that testing the conjectured bound requires computing the behavior of the ZDMs as well.

\section{Alternative critical points and the appearance of zero-damped modes}
\label{sec:ZDMs}

\subsection{Choices of $r_0$}

We now turn to a subtlety that we have thus far swept under the rug. The algorithm we presented in  \S\ref{sec:algorithm} computes  a perturbation series around a point $r_0$ that is determined by the equation \eqref{eq:r0extrema}. When either the scalar field $\phi$ or the black hole is uncharged, these points  are  simply extrema  of the potential $V(r)$, as in the original Bender--Wu calculation. In general, however, $r_0$ depends not only on $V(r)$, but on the gauge potential $K(r)$. In all of the calculations we have presented in \S\ref{sec:validateDMs}, we have been perturbing around the smallest real $r_0 > r_+$ for which \eqref{eq:r0extrema} is satisfied. To apply the method, it is important to know whether multiple choices of $r_0$ will lead to correct calculations of quasinormal mode eigenvalues.

\begin{figure}[!h]
\centering
\includegraphics[width=0.55\textwidth]{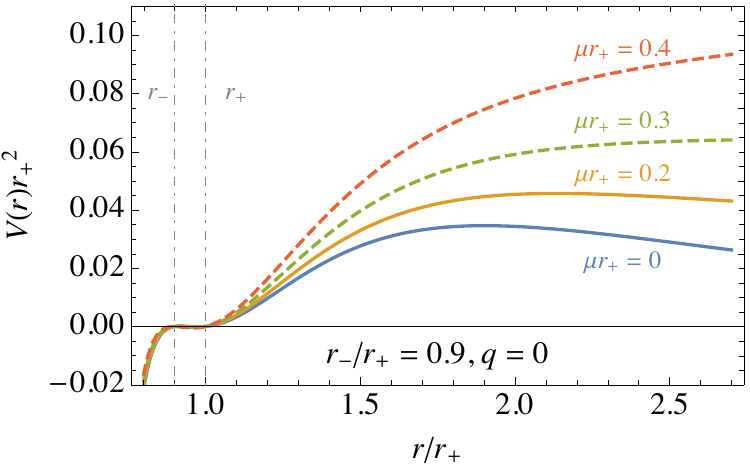}
\caption{Dependence of the potential $V(r)$ for a massive, neutral scalar field in a $D=4$ Reissner--Nordstr{\"o}m background on the  scalar mass parameter $\mu$. We have fixed the ratio of inner and outer horizon coordinates, $r_- / r_+ = 0.9$. The potential has a maximum at $r_0 > r_+$ when  $\mu r_+ \lesssim 0.29$, as in the solid blue and orange curves. For larger values of $\mu$, as in the dashed green  and red  curves, there is no maximum at $r > r_+$. For all values of $\mu r_+$, there is a minimum and a maximum of $V(r)$ behind the outer horizon.}
\label{fig:Vmudep}
\end{figure} 

To illustrate the physics, let us focus for the moment on the case of a neutral scalar, $q = 0$, for which choices of $r_0$ correspond to critical points of the potential. We plot the potential $V(r)$  in Fig.~\ref{fig:Vmudep} for several choices of $\mu$ at fixed $r_\pm$. We see that for sufficiently small $\mu$, as in all of our plots in \S\ref{sec:validateDMs}, there is a {\em maximum} of the potential at a choice of $r_0 > r_+$. This fits well with the motivation of Hatsuda's method that we discussed in \S\ref{subsec:Hatsudaalgorithm}: when analytically continuing our quasinormal mode problem to a Schr{\"o}dinger problem, the potential is inverted, so the maximum of $V(r)$ becomes a quadratic minimum in which we can find standard quantum mechanical bound states. Fig.~\ref{fig:Vmudep} shows that above a critical $\mu = \mu_\mathrm{c}$, this maximum of $V(r)$ disappears. This raises the question: can the method be applied at all when $\mu > \mu_\mathrm{c}$? We also see that there is a second maximum of the potential, at {\em small} $r$, inside the black hole horizon. This maximum persists at larger $\mu$. Does perturbing around this other maximum allow us to compute physically interesting modes? Here we present some tentative, preliminary steps toward addressing such questions. We do not yet have complete answers. In particular, our discussion in \S\ref{subsec:bcs} of why we expect our results to obey the correct quasinormal mode boundary conditions applies when perturbing around real $r_0 > r_+$, and does not extend in an obvious way to the more general results found in this section. Nonetheless, empirically, we will see that the algorithm produces correct quasinormal mode frequencies in more general cases.

\subsection{Extrapolating to larger $\mu$}
\label{subsec:largermu}

\begin{figure}[!h]
\centering
\includegraphics[width=0.55\textwidth]{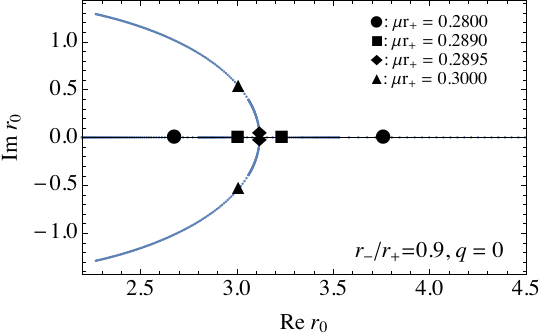}
\caption{Dependence of the critical points, $r_0$, of the potential $V(r)$ on the scalar mass parameter $\mu r_+$, for the fixed choice $r_-/r_+ = 0.9$. Initially, there is a real maximum of $V(r)$ on the left and a  real minimum on the right. As $\mu \to \mu_\mathrm{c}$, these two critical points merge and then move away from the real axis, appearing as two complex conjugate critical points. Some specific choices of $\mu r_+$, labeled in the plot, are marked to illustrate the behavior.}
\label{fig:r0valuesRm09}
\end{figure}

We have seen that the maximum of $V(r)$ is only present below a critical mass $\mu_\mathrm{c}$. The potential $V(r)$ is an analytic function of $r$ away from the origin, and is analytic in $\mu^2$. Hence, a critical point cannot simply ``disappear'' when we cross $\mu_\mathrm{c}$. Rather, what happens is that two real critical points, a maximum and a minimum of $V(r)$, merge at $\mu_\mathrm{c}$ and then, when $\mu > \mu_\mathrm{c}$, separate and move away from the real axis. This behavior is illustrated in Fig.~\ref{fig:r0valuesRm09}. The complex values of $r_0$ appear as complex conjugate pairs. As we mentioned in \S\ref{sec:algorithm}, our algorithm involves various sign choices. We only present numerical results for $\beta < 0, s_0 = 1$, but the frequencies computed at one value of $r_0$ are related to those computed at the complex conjugate $r_0^*$ through appropriate changes of signs, as in \eqref{eq:signchoices}.

\begin{figure}[!h]
\centering
\includegraphics[width=\textwidth]{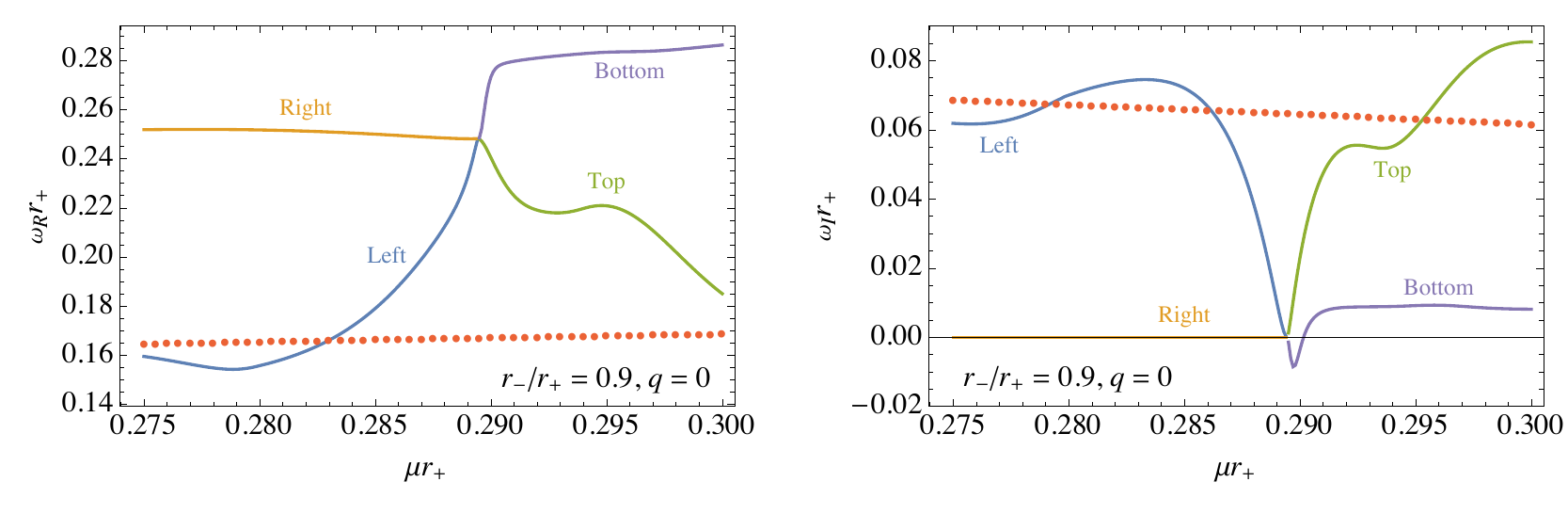}
\caption{Breakdown of the algorithm near $\mu = \mu_\mathrm{c}$, where the critical points $r_0$ bifurcate as shown in Fig.~\ref{fig:r0valuesRm09}. As in the plots of \S\ref{sec:validateDMs}, the red dots show results we obtained with Leaver's method, which we believe to be an accurate calculation of the quasinormal mode frequency. This result shows that Hatsuda's algorithm breaks down near the critical $\mu$ where a bifurcation in the critical points of $V(r)$ occurs. None of the branches of $r_0$ accurately match the true quasinormal modes in this region.}
\label{fig:mucriticalRm09}
\end{figure} 

\begin{figure}[!h]
\centering
\includegraphics[width=\textwidth]{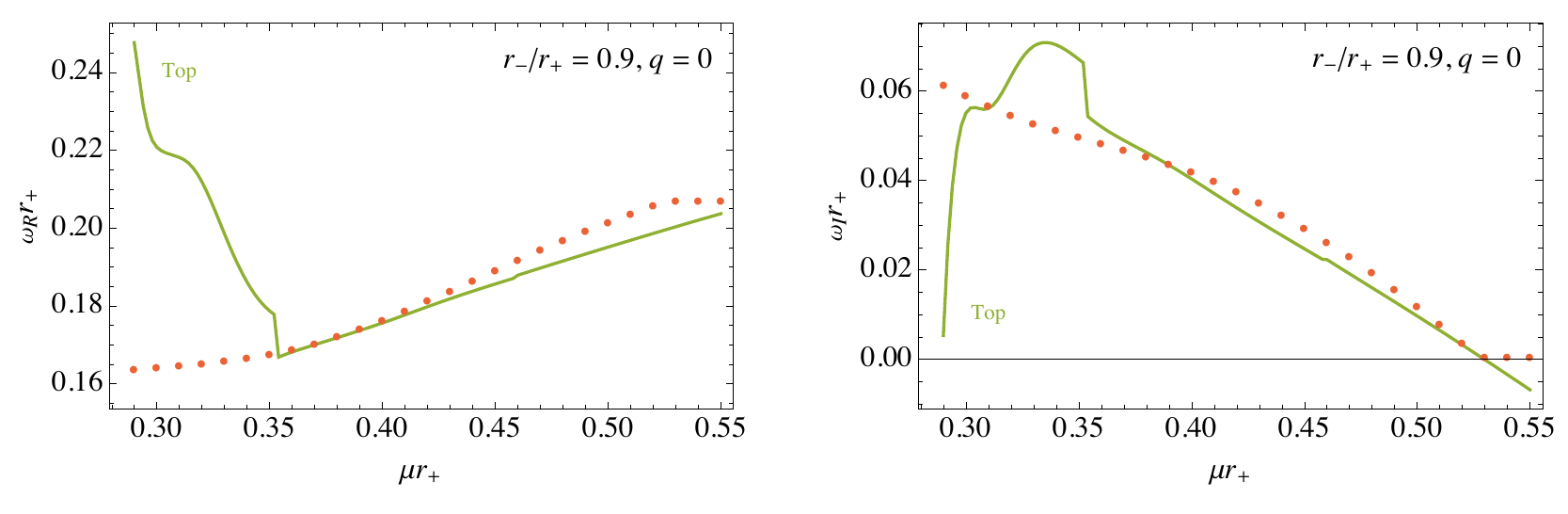}
\caption{Calculation of the quasinormal mode frequencies for $\mu > \mu_\mathrm{c}$ using the branch of $r_0$ labeled ``top'' in Fig.~\ref{fig:mucriticalRm09} (continuous green curve) as well as Leaver's method (red dots). Although the method breaks down badly near $\mu_\mathrm{c}$, we see that it approximately reproduces the quasinormal mode with smallest $\omI$ above $\mu_\mathrm{c}$ until this mode disappears at $\mu r_+ \approx 0.53$.}
\label{fig:muabovecriticalRm09}
\end{figure} 

To explore how our algorithm behaves in general, we have seeded it with both of the critical points with  ${\rm Re}\, r_0 > r_+$  in both the $\mu < \mu_\mathrm{c}$ and $\mu > \mu_\mathrm{c}$ regimes. We label the critical points in the former case ``left'' and ``right'' and in the latter ``top'' and ``bottom.'' The results are shown in Fig.~\ref{fig:mucriticalRm09}, along with the mode of smallest $\omI$ as determined by Leaver's method. We see that the values computed with Hatsuda's method for the left and right branches merge at $\mu = \mu_\mathrm{c}$ and then split apart into the top and bottom branches, but the calculations significantly disagree with the results of Leaver's method. We interpret this to mean that our algorithm breaks down badly in the vicinity of the bifurcation point, when there are two nearby critical points in $r_0$, and should not be trusted in this regime. We see from Fig.~\ref{fig:mucriticalRm09} that at larger $\mu$ values beyond $\mu_\mathrm{c}$, the branch labeled ``top'' begins to approach the solution obtained with Leaver's method. In Fig.~\ref{fig:muabovecriticalRm09}, we continue our calculation to larger $\mu$, and find that the top branch indeed comes close to the computation with Leaver's method once $\mu r_+ \gtrsim 0.35$. This particular mode has a decreasing $\omI$ which reaches zero at $\mu r_+ \approx 0.53$. This phenomenon was first observed in \cite{Ohashi:2004wr}, which referred to the modes at the point where $\omI$ vanishes as quasi-resonant modes (QRMs). These modes were later studied in detail,  including in cases with rotating or higher-dimensional black holes \cite{Konoplya:2004wg,Konoplya:2006br,Zhidenko:2006rs}. Beyond this value of $\mu r_+$, this quasinormal mode no longer exists. In Leaver's method, we find that the value of $\omega$ crosses through a branch cut in the continued-fraction expression whose roots we are solving for. The values returned for $\mu r_+ \gtrsim 0.53$ have zero imaginary part and are not to be trusted as physical modes. Hatsuda's method continues smoothly past $\mu r_+ \gtrsim 0.53$, but returns results with negative values of $\omI$, which again should not be interpreted as physical quasinormal modes. We find it intriguing that there is a regime of $\mu$ over which Hatsuda's algorithm applied to the upper complex value of $r_0$ returns answers in good agreement with Leaver's method, as this suggests that the algorithm has validity even for complex $r_0$. On the other hand, we have attempted to compute excited modes of higher $\nu$ (and larger $\omI$), which would survive beyond $\mu r_+ \approx 0.53$. We can find such modes with Leaver's method, but  we find that the values obtained with Hatsuda's algorithm do not converge well and have large uncertainty. This does not necessarily mean that the algorithm cannot find multiple quasinormal modes when expanding around complex choices of $r_0$, but at least for these particular parameters, the convergence is slow enough that we have not  yet obtained reliable results.

\subsection{ZDMs of scalar fields in the Reissner--Nordstr{\"o}m background}

As we have remarked above, our earlier results in \S\ref{sec:validateDMs} are all for damped modes, i.e., modes which have $\omI$ nonzero even in the extremal limit. A natural question is whether our method can also calculate the frequencies of zero-damped modes. One numerical study of quasinormal modes of charged, massive scalar fields in the background of charged black holes that showed branching into two families near extremality was \cite{Konoplya:2013rxa}. A more explicit numerical demonstration of a crossover between DMs and ZDMs near extremality appeared in \cite{Richartz:2014jla}. ZDMs of scalar fields in near-extremal Reissner--Nordstr{\"o}m backgrounds have also been found analytically \cite{Hod:2010hw,Hod:2017uqc}, as have a closely related set of modes in a charged AdS--Rindler background \cite{Urbano:2018kax}.

\begin{figure}[!h]
\centering
\includegraphics[width=\textwidth]{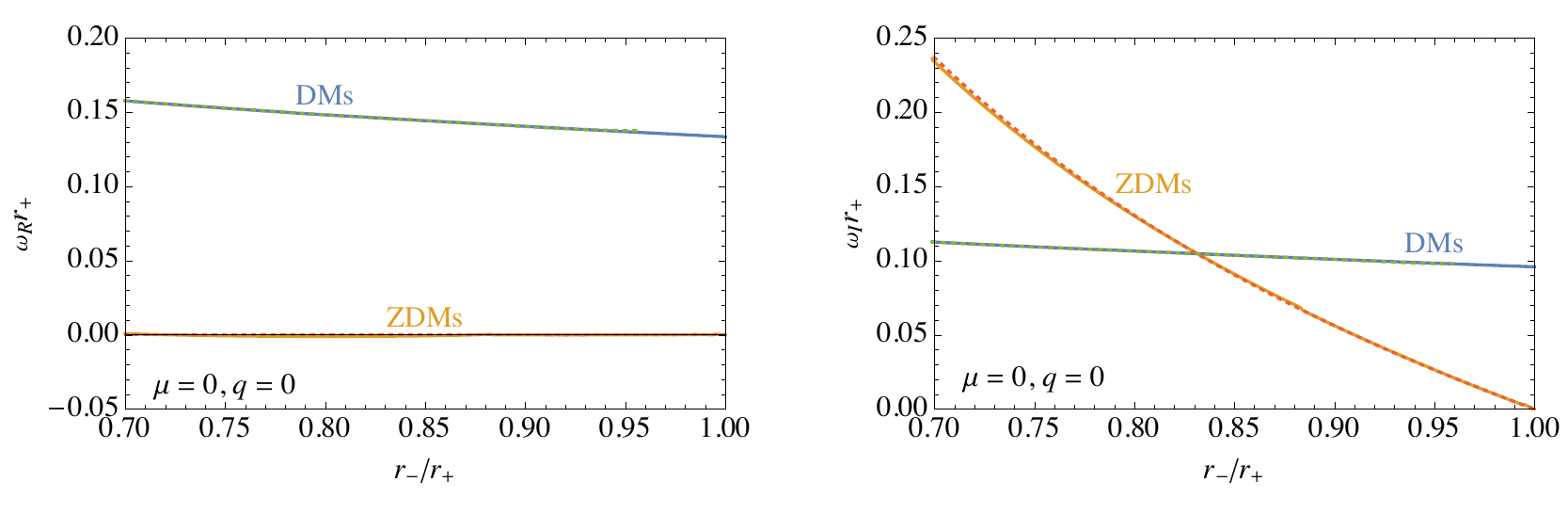}
\caption{Quasinormal modes of the massless, neutral scalar field in Reissner--Nordstr{\"o}m backgrounds near extremality. Solid blue and orange curves are DMs and ZDMs computed with Hatsuda's algorithm. The ZDMs are computed by expanding around the maximum of $V(r)$ behind the outer horizon. Dotted green and red curves are the same modes computed with Leaver's method.}
\label{fig:mu0q0DMZDM}
\end{figure} 

\begin{figure}[!h]
\centering
\includegraphics[width=0.55\textwidth]{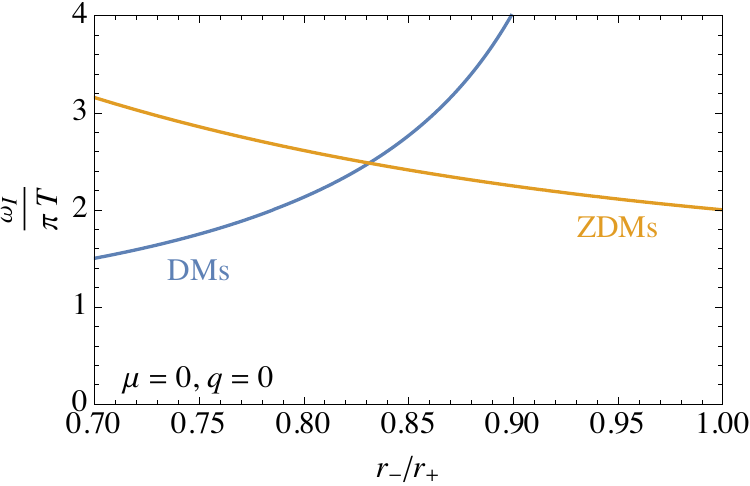}
\caption{Imaginary part $\omI$ of the quasinormal modes of the massless, neutral scalar field in Reissner--Nordstr{\"o}m backgrounds near extremality, plotted in units of the black hole temperature. This is the same data as the right-hand panel of Fig.~\ref{fig:mu0q0DMZDM}, but shows more clearly that $\omI \propto T$ in the extremal limit for the ZDMs.}
\label{fig:piTdepmu0q0DMZDM}
\end{figure}

Since our algorithm is most well behaved when perturbing around a real maximum of $V(r)$, a natural step is to calculate the modes of a neutral, massless scalar by choosing the maximum of the potential at $r_0 < r_+$ that is visible in Fig.~\ref{fig:Vmudep}. We find that our computation is well behaved in this limit and that the $\nu  = 0$ mode matches the ZDM with smallest $\omI$ that we can find with Leaver's method. This result is illustrated in Fig.~\ref{fig:mu0q0DMZDM}, which shows both the DMs and ZDMs. The mode  of smallest $\omI$ switches from a damped mode to a ZDM at $r_- / r_+ \approx 0.83$, which is very close to extremality: $Q/Q_\mathrm{ext} \approx 0.996$. This is quite similar to the result for quasinormal modes of Dirac charged particles presented in Fig.~4 of \cite{Richartz:2014jla}. For the ZDMs, $\omI$ approaches zero linearly with temperature in the near-extremal limit. We illustrate this in Fig.~\ref{fig:piTdepmu0q0DMZDM} by plotting $\omI/(\cpi T)$, which approaches $2$ as $r_- \to r_+$. Although we have not provided an argument for why expanding around $r_0 < r_+$ computes modes with the correct quasinormal mode boundary conditions, as opposed to some other set of solutions, the empirical agreement with Leaver's method gives us confidence that the algorithm works in this regime.

\begin{figure}[!h]
\centering
\includegraphics[width=0.55\textwidth]{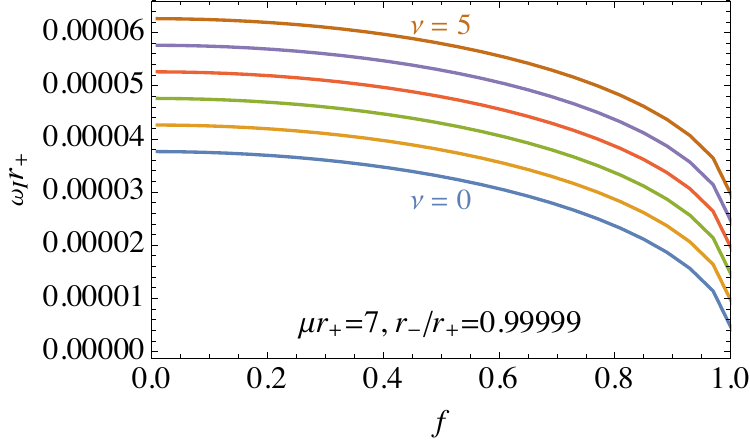}
\caption{Imaginary part $\omI$ of the $l = 0$ ZDMs of a massive, charged scalar field in a nearly-extremal Reissner--Nordstr{\"o}m background, computed with our extension of Hatsuda's method when expanding around a critical point $r_0$ inside the outer horizon. We have fixed the mass parameter $\mu r_+ = 7$, and vary the scalar charge parameter $f = \sqrt{2}q/(\kappa \mu)$. We plot the six lowest-lying ZDMs, with $\nu$ increasing from $0$ for the lowest curve to $5$ for the highest curve. These results agree to high precision with analytic results of Hod \cite{Hod:2010hw}.}
\label{fig:chargedZDMs}
\end{figure}

\begin{figure}[!h]
\centering
\includegraphics[width=0.55\textwidth]{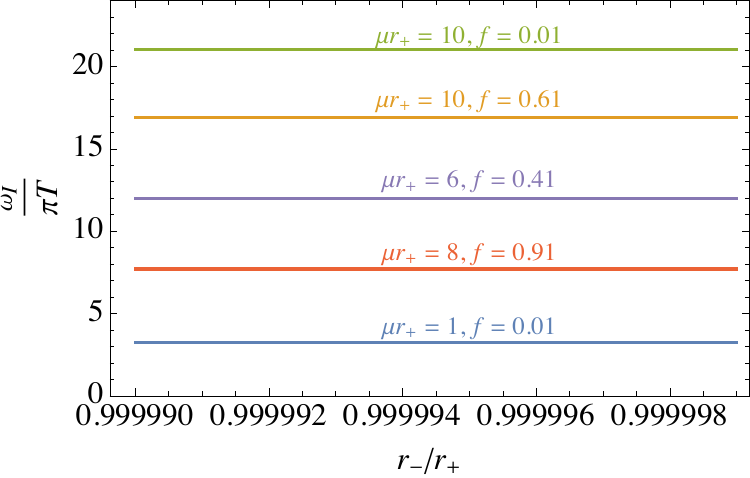}
\caption{The ratio of $\omI$ to $\cpi T$ for $\nu = 0, l = 0$ ZDMs of a massive, charged scalar field in a very near-extremal Reissner--Nordstr{\"o}m background, for selected choices of $\mu r_+$ and $f$ (labeled curves). We see that to high precision, the ZDMs computed with our method have $\omI$ approaching a constant ($\mu$- and $q$-dependent) multiple of temperature in the extremal limit.}
\label{fig:chargedZDMsPiT}
\end{figure}

When we turn on a charge $q$ for the scalar field, critical points (solutions $r_0$ to equation \eqref{eq:r0extrema}) with $\mathrm{Re}\, r_0 < r_+$ continue to exist, though in general they develop an imaginary part. As we discussed in the previous subsection, we do not fully understand how our algorithm behaves for complex $r_0$, though for certain choices it appears to compute correct quasinormal frequencies. Motivated by this, we have empirically found that there is a choice of $r_0$ behind the outer horizon with small negative imaginary part, which for small $f$ and $r_- \approx r_+$, is approximately given by
\begin{equation}
r_0 \approx \frac{r_+ + r_-}{2} - \iu \frac{(r_+ -  r_-)  f}{2}.
\end{equation}
Expanding around this point, we obtain reasonable quasinormal mode frequencies, which are ZDMs. For example, in Fig.~\ref{fig:chargedZDMs}, we have fixed a large mass $\mu r_+ = 7$ and varied the charge fraction $f$. This choice of $\mu$ is larger than $\mu_\mathrm{c}$, but because we are perturbing around the critical point {\em inside} the outer horizon, this poses no difficulty for the algorithm. We have checked that these results agree to high precision with the formula of \cite{Hod:2010hw}, which when converted to the conventions we introduced in \S\ref{sec:minireview} is as follows:
\begin{equation}
\omega = \frac{q Q}{2} T - 2\cpi \iu T \left[\nu + \frac{1}{2} + \sqrt{\frac{1}{4} + l(l+1) + r_Q^2 \mu^2 - \frac{(qQ)^2}{16\cpi^2}}\right].
\end{equation}
This is further illustrated in Fig.~\ref{fig:chargedZDMsPiT}, which shows the ratio of $\omI$ to $\cpi T$ for various ZDMs calculated very close to extremality. We find that $\omI$ goes to zero linearly with temperature in the extremal limit in precisely the manner predicted by the analytic formula. Hence, we have convincing empirical evidence that perturbing around a complex critical point $r_0$ behind the horizon correctly computes the quasinormal frequencies of zero-damped modes, for a range of $\mu$ and $f > 0$. We have encountered numerical difficulties when attempting to extrapolate these results to modes of higher $l$ or $f < 0$.

\section{Conclusions and outlook}
\label{sec:outlook}

In this paper, we have presented a numerical method for solving for quasinormal modes of charged fields in the background of a charged black hole. This extends the recent work of Hatsuda \cite{Hatsuda:2019eoj}. We have found that this method can not only compute damped quasinormal modes, but also, at least in some cases, the zero-damped modes for which $\omI \to 0$ as $T \to 0$. One appealing feature of this method is that it is requires little analytic work to apply. Given the functions $K(r)$ and $V(r)$, one simply has to numerically find a candidate critical radius $r_0$ from \eqref{eq:r0extrema} and then run the procedure. This requires less overhead than setting up, for example, Leaver's method for a new class of quasinormal mode problems. Another interesting feature of our approach is that different families of modes, for instance DMs and ZDMs, seem to be computed around different critical values of $r_0$, so that they are easily separated. On the other hand, a disadvantage of our method is that it breaks down near bifurcation points in $r_0$, as shown in \S\ref{subsec:largermu}. In some cases we have also encountered numerical difficulties when attempting to compute higher modes at a complex $r_0$. We do not yet have a detailed analytic understanding of how all of the sign choices in the algorithm are related to boundary conditions, especially when perturbing around values $r_0 < r_+$ or complex $r_0$, and it is possible that an improved understanding could allow the algorithm to be used more effectively. In the simplest case of perturbing around real $r_0 > r_+$, we have provided an argument (but not a rigorous proof) in \S\ref{subsec:bcs} that the algorithm finds modes obeying the correct quasinormal mode boundary conditions, closely following a similar argument in \cite{Hatsuda:2019eoj}.

As we mentioned in the introduction, the existence of ZDMs in near-extremal limits is crucial for theories to satisfy Hod's conjectured ``universal relaxation bound'' $\omI \lesssim \cpi T_{\rm BH}$  \cite{Hod:2006jw, Hod:2007tb, Hod:2008zz, Gruzinov:2007ai, Hod:2010hw}. This is an interesting conjecture in part because it connects to topics of interest in other corners of physics: the study of ``Planckian metals'' in condensed matter physics (so called because they have a dissipation or transport rate $\tau \sim \hbar/T$) \cite{bruin2013similarity,legros2019universal, cao2019strange, nakajima2019planckian, Patel:2019qce} and the Weak Gravity Conjecture (WGC) \cite{ArkaniHamed:2006dz, Heidenreich:2015nta} in quantum gravity. The latter connection is discussed in \cite{Hod:2017uqc, Urbano:2018kax} by studying the quasinormal modes of charged fields in the near-extremal limit. To assess the extent to which the universal relaxation bound is related to the  WGC, one should extend these calculations away from the strict extremal limit, understand whether the ZDMs of gravitational and electromagnetic perturbations obey the bound or not, and also explore calculations in a more general number of spacetime dimensions $D > 4$. For gravitational and electromagnetic perturbations, studies in $D > 4$ should make use of the results of Kodama and Ishibashi \cite{Kodama:2003kk}. The effective potentials are more complicated than in the $D = 4$ case. We will present results of such studies in a subsequent publication.

\section*{Acknowledgments}

We would especially like to thank Yasuyuki Hatsuda for sharing his Mathematica code for computing quasinormal modes using the algorithm of \cite{Hatsuda:2019eoj}. We thank Patrick Draper, Ben Heidenreich, and Tom Rudelius for useful discussions. DSE was supported by a Harvard University Program for Research in Science and Engineering (PRISE) Fellowship. MR is supported in part by the DOE Grant DE-SC0013607.

\bibliography{ref}
\bibliographystyle{utphys}

\end{document}